\documentclass[twocolumn,showpacs,preprintnumbers,amsmath,amssymb,superscriptaddress,floatfix]{revtex4}
\usepackage{graphicx}
\usepackage{amsmath}
\usepackage{bm}
\usepackage{mathrsfs}
\usepackage{color}
\usepackage{slashed}
\usepackage{dcolumn}

\newcommand{\bald}[1]{{\bf #1}}

\newcommand{\eqf}[1]{\begin{equation}\begin{split}#1\end{split}\end{equation}}

\newcommand{{\vp}}{{\vec p}}
\newcommand{{\vq}}{{\vec q}}

\newcommand{\beq}{\begin{equation}}
\newcommand{\eeq}[1]{\label{#1} \end{equation}}

\newcommand{\lton}{\mathrel{\lower.9ex
                  \hbox{$\stackrel{\displaystyle <}{\sim}$}}}
\newcommand{\ee}{\end{equation}} \newcommand{\ben}{\begin{enumerate}}
\newcommand{\een}{\end{enumerate}} \newcommand{\bit}{\begin{itemize}}
\newcommand{\eit}{\end{itemize}} \newcommand{\bc}{\begin{center}}
\newcommand{\ec}{\end{center}} \newcommand{\bea}{\begin{eqnarray}}
\newcommand{\eea}{\end{eqnarray}}
\newcommand{\beqar}{\begin{eqnarray}}
\newcommand{\eeqar}[1]{\label{#1} \end{eqnarray}}

\begin{document}

\title{ The physics of $Z^0/\gamma^*$-tagged jets at the LHC }

\author{R. B. Neufeld}
  \email{neufeld@lanl.gov}
\affiliation{Los Alamos National Laboratory, Theoretical Division, MS B238,
Los Alamos, NM 87545, U.S.A.}
 \author{Ivan Vitev}
   \email{ivitev@lanl.gov}
\affiliation{Los Alamos National Laboratory, Theoretical Division, MS B238,
Los Alamos, NM 87545, U.S.A.}
\author{Ben-Wei Zhang}
 \email{bwzhang@iopp.ccnu.edu.cn}
  \affiliation{Key Laboratory of Quark and Lepton Physics (Central China Normal University),
Ministry of Education, People's Republic of China}
   \affiliation{Los Alamos National Laboratory, Theoretical Division, MS B238,
Los Alamos, NM 87545, U.S.A.}

\date{\today}

\begin{abstract}
Electroweak bosons produced in conjunction with jets in high-energy collider experiments is one of the principle final-state channels that can be used to test the accuracy of perturbative Quantum Chromodynamics calculations and to assess the potential to uncover new physics through comparison between data and theory.  In this paper we present results for the $Z^0/\gamma^*$+jet production cross sections at the LHC at leading and next-to-leading orders. In proton-proton reactions we elucidate up to ${\cal O}(G_F\alpha_s^2)$ the constraints that jet tagging via the $Z^0/\gamma^*$ decay dileptons provides on the momentum distribution of jets. In nucleus-nucleus reactions we demonstrate that tagged jets can probe important aspects of the dynamics of quark and gluon propagation in hot and dense nuclear matter and characterize the properties of the medium-induced parton showers in ways not possible with more inclusive measurements. Finally, we present specific predictions for the anticipated suppression of the $Z^0/\gamma^*$+jet production cross section in the quark-gluon plasma that is expected to be created in central lead-lead collisions at the LHC relative to the naive superposition of independent nucleon-nucleon scatterings.
\end{abstract}


\maketitle

\section{Introduction}

Hadronic jets~\cite{Sterman:1977wj} produced in today's high-energy collider experiments have long been regarded as a premier tool to test the fundamentals of perturbative ($Q^2 \gg \Lambda_{QCD}^2$) Quantum Chromodynamics (QCD)~\cite{eks}. The start-up of the Large Hadron Collider (LHC) has stimulated new theoretical and experimental interest in jet
observables~\cite{Campbell:2006wx}. At the LHC, final states involving high jet multiplicities, jets+long-lived heavy leptons and jets+missing energy are among the most-promising channels for discovery of physics beyond the Standard Model. It is, therefore, critical to understand theoretically as accurately as possible the QCD background to signatures of new physics.

One of the most studied QCD final states at collider energies is an electroweak boson accompanied by jets~\cite{Kajantie:1978qv,Campbell:2005zv,Campbell:2003dd,Campbell:2002tg}. Of these processes, the $Z^0/\gamma^*$+jet production is relatively easy to measure via the  $Z^0/\gamma^*\rightarrow l^++l^-$ decay channel. Comparison between
theory and experiment has so far only been carried out at the Tevatron $\sqrt{s}=1.96$~TeV~\cite{Abazov:2008ez} without restrictions on the momentum of the vector boson. In this paper we elucidate the constraints on the production and momentum distribution of jets associated with $Z^0/\gamma^*$ of {\em fixed} $p_T$ to lowest [${\cal O}$($G_F\alpha_s$)] and next-to-leading [${\cal O}$($G_F\alpha_s^2$)] orders~\cite{Campbell:2002tg}. The cross sections that we find are also the much needed baseline for the investigation of many-body QCD effects in heavy ion reactions at the LHC.

\begin{figure*}
\centerline{
\includegraphics[width = 0.43\linewidth]{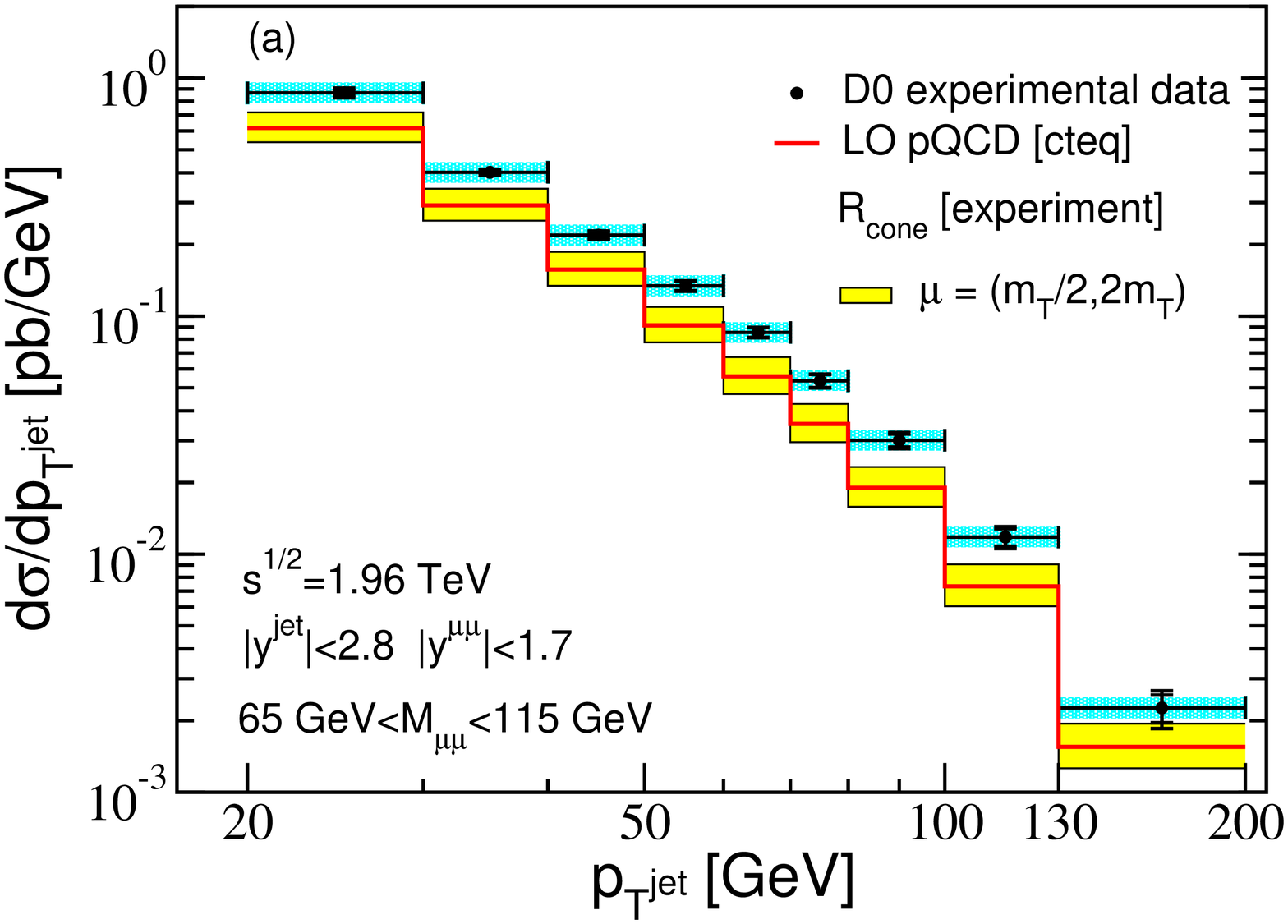}\hskip0.04\linewidth
\includegraphics[width = 0.43\linewidth]{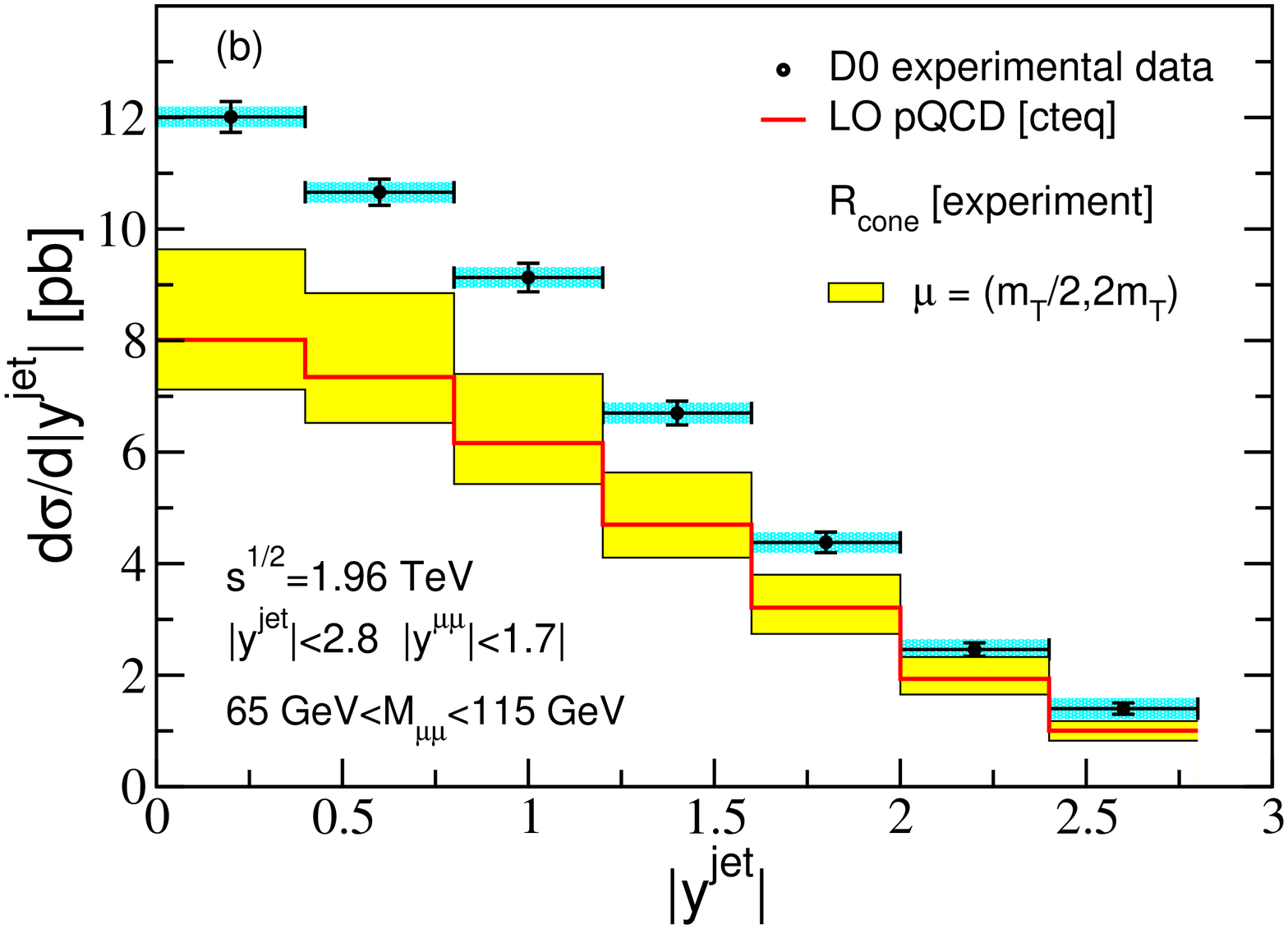}
}
\caption{(Color online) Experimental results from the Fermi Lab Tevatron Collider \cite{Abazov:2008ez} for the cross section of jets associated with $Z^0$/$\gamma^*\rightarrow \mu^++\mu^-$ in $p+\bar{p}$ collisions at $\sqrt{s} = 1.96$ TeV are compared to tree level pQCD. The left panel shows this cross section versus  $p_{T\,(\rm jet)}$, whereas the right panel shows it versus $y_{(\rm jet)}$. The (red) lines indicate our LO result. The (yellow) bands indicate the variation from the choice of scale. The experimental cuts are discussed in the text.}
\label{ourdata}
\end{figure*}

A significant part of the LHC research program involves reactions with ultrarelativistic nuclei - specifically, Pb+Pb (lead-lead) collisions up to a center-of-mass energy per nucleon pair $\sqrt{s_{NN}}=5.5$~TeV. These heavy ion runs will consolidate the evidence for the creation of a deconfined state-of-matter at extremely high temperature, the quark-gluon plasma (QGP), and probe its properties in a new energy regime~\cite{BraunMunzinger:2007zz}. One of the principal advantages of the LHC over the Relativistic Heavy Ion Collider (RHIC) is that the projected 28-fold increase in energy will open final-state channels for jet tomography of the QGP that are currently inaccessible to experiment or limited by statistics~\cite{Vitev:2008rz}.  One such channel is $Z^0/\gamma^*$+jet, which has been previously discussed in the context of the energy or momentum constraints that tagging of the decay
dileptons can provide on the away-side parton shower~\cite{Srivastava:2002kg,Kunde:1900zz,Lokhtin:2004zb}. The short $Z^0$ production time, $\tau_{prod.} \sim 1/m_T^Z$, and subsequent decay, $\tau_{decay} \sim 1/\Gamma$, imply that these processes occur before or during the formation of the QGP. While the produced dileptons have to traverse a region of dense nuclear matter, they reach the detectors unscathed by the strong interactions in the deconfined QCD medium.

The utility of jet tagging, however, extends well beyond the simple determination of the energy of the parent parton. It allows one to probe aspects of the QCD many-body dynamics in nucleus-nucleus (A+A) collisions that are inaccessible via leading particles or even leading particle correlation measurements~\cite{Vitev:2009rd}. One of the most striking results from the heavy-ion program at RHIC~\cite{Gyulassy:2003mc,Jacobs:2004qv} is the attenuation of the flux of energetic hadrons in the plasma, known as jet quenching. Much theoretical effort has been invested in understanding the mechanism of this suppression and, in particular, of partonic energy loss in the QGP~\cite{Baier:1996sk,Zakharov:1997uu,Gyulassy:2000fs,Guo:2000nz}. Unfortunately, leading particle quenching alone~\cite{Sharma:2009hn} is not sufficient to discriminate between partonic energy loss formalisms or to extract quantitatively the stopping power of the QGP for color-charged particles~\cite{Bass:2008rv}. In contrast, jet observables are much more closely related to the underlying perturbative QCD theory and to the characteristics of the vacuum and medium-modified parton showers. Hence, they are also much more discriminating with respect to theoretical approximations and model assumptions.

In this paper, we extend the formalism developed to evaluate inclusive jet cross sections and shapes in A+A reactions at lowest and next-to-leading orders~\cite{Vitev:2008rz,Vitev:2009rd} to  tagged jets. We present the results of a systematic study of the production and subsequent suppression of the $Z^0/\gamma^*$+jet final state in central Pb+Pb collisions at the LHC at $\sqrt{s_{NN}}=4$~TeV to ${\cal O}$($G_F\alpha_s^2$). We demonstrate that tagged jets can provide insight into the multiplicity and distribution of gluons induced by parton propagation in the QGP. Much of the discrepancy in the determination of the QGP properties can be traced to differences in the characteristics of these distributions~\cite{Baier:2001yt,Gyulassy:2001nm}.  In our complete numerical examples for the modification of the $Z^0/\gamma^*$ tagged jet cross section we use the Gyulassy-Levai-Vitev (GLV) approach to the non-Abelian energy loss of quarks and gluons propagating in dense nuclear matter~\cite{Vitev:2007ve}.

Our work is organized as follows: in section II we present results for the cross section of the $Z^0/\gamma^*+{\rm jet}$ channel at the Tevatron and at the LHC. We elucidate the crucial differences between the lowest-order and next-to-leading order calculations for the $p_T$-tagged jet cross section where the transverse momentum of the boson is constrained in a narrow interval via its decay products ($l^++l^-$). In section III we discuss one of the principle differences between proton-proton and nucleus-nucleus collisions: the radiative corrections from final-state interactions of $Z^0/\gamma^*$-tagged quark and gluon jets in the QGP. We give details for the calculation of the medium induced bremsstrahlung at LHC energies. Our results for the suppression of the tagged jet cross sections in central lead-lead (Pb+Pb) collisions are given in section~IV. We demonstrate how tagged cross sections can probe essential aspects of the physics of parton energy loss in strongly-interacting matter in ways not possible with inclusive jets or leading particles. Our conclusions are given in section V. Appendix A contains selected steps in the calculation of the lowest order tagged cross sections. In appendix B we briefly discuss the Dalitz decays of the $Z^0/\gamma^*$ to dileptons. Finally, the sensitivity of the experimentally observed cross section suppression to the fraction of quark- and gluon-initiated jets is discussed in Appendix C.

\section{$Z^0$/$\gamma^*$-tagged jets in hadronic collisions}
\label{ppcross}

\begin{figure*}
\vskip0.04\linewidth
\centerline{
\includegraphics[width = 0.43\linewidth]{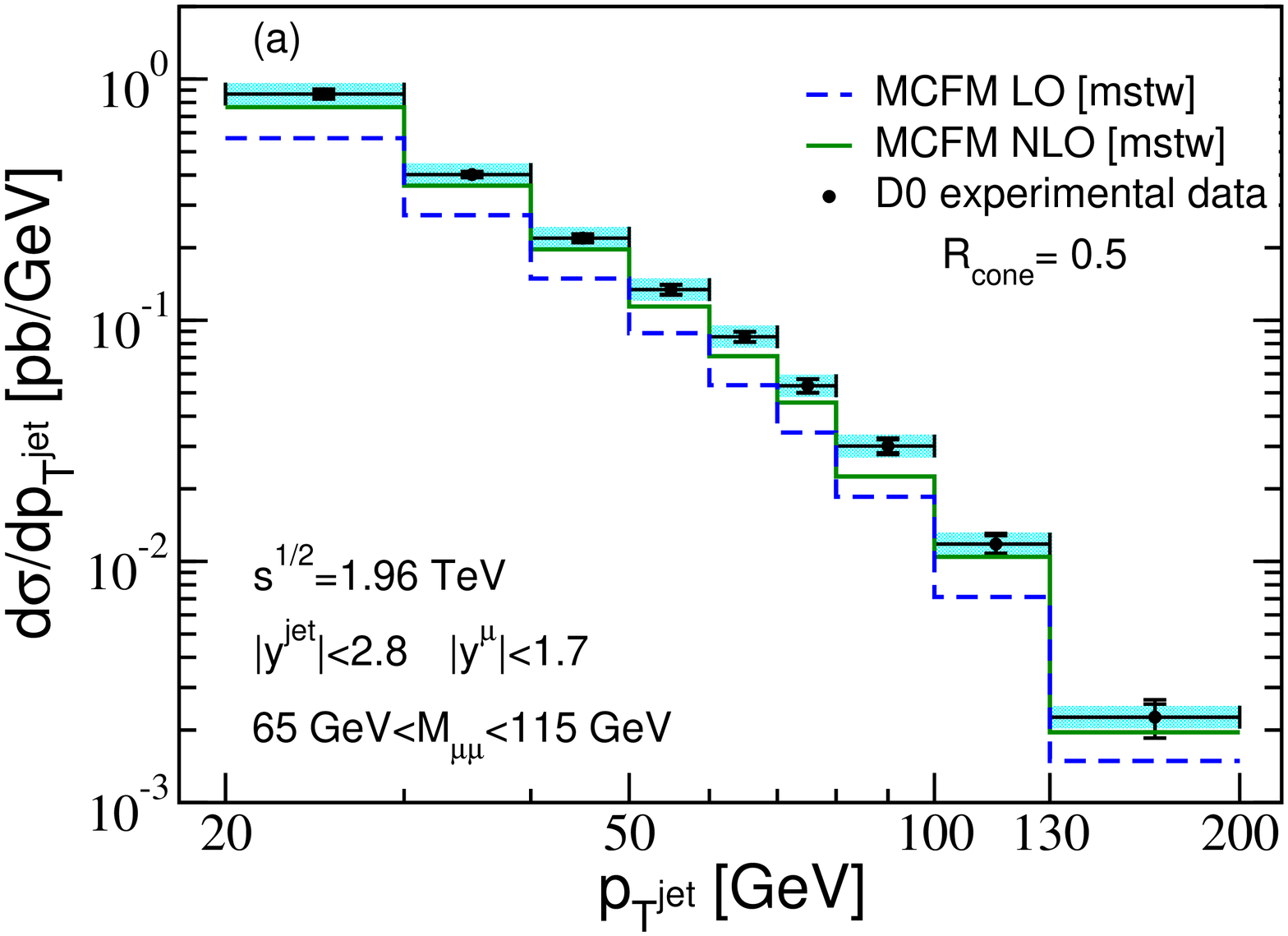}\hskip0.04\linewidth
\includegraphics[width = 0.43\linewidth]{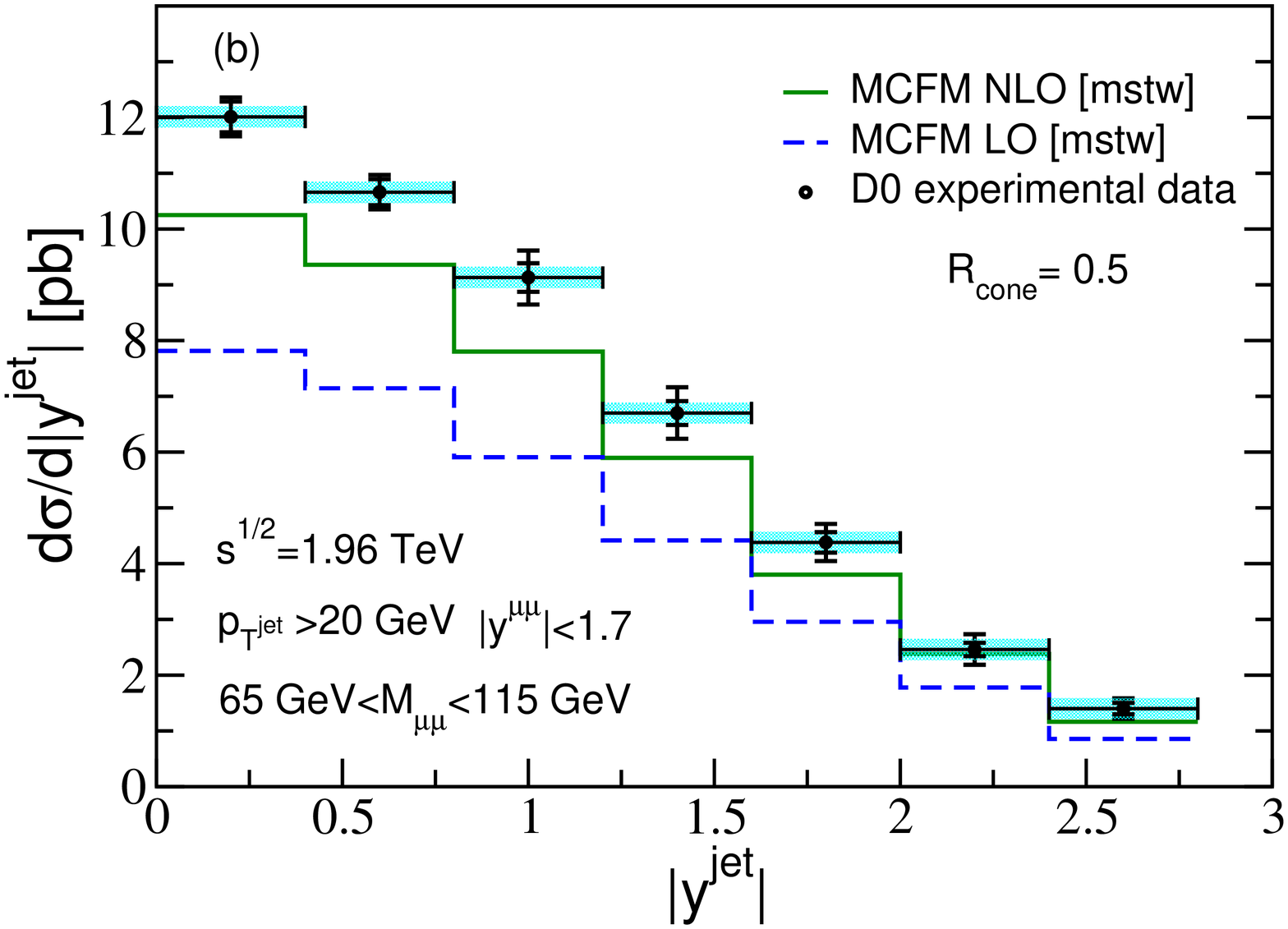}
}
\caption{(Color online) Same as in Figure \ref{tevdata} but with a comparison of the Tevatron data to the MCFM LO result (blue) dotted line and the MCFM NLO result (green) solid line. Note that experimental error bars are statistical only.}
\label{tevdata}
\end{figure*}

We begin by discussing the cross section for $Z^0/\gamma^*$-tagged jet production in hadron-hadron collisions.  It is instructive to first consider the leading order (LO) result, from which one can understand the underlying production processes and appreciate why the $Z^0$ boson was originally considered as a  suitable tag for the initial associated jet energy~\cite{Srivastava:2002kg}. Details of this calculation are given in appendix~\ref{lo}.

In the collinear factorization approach, the $Z^0/\gamma^*$+jet cross section reads:
\begin{eqnarray}
&&\frac{d \sigma}{d y_{(Z)} \, d y_{\rm(jet)} \, d^2 p_{T\,(Z)} d^2 p_{T\,\rm (jet)} }  =  \sum_{g,q,\bar{q}} f(\bar{x}_1,\mu)f(\bar{x}_2,\mu)
\nonumber \\
 && \times  \frac{|M|^2}{(2\pi)^2 \, 4 \, \bar{x}_1 \, \bar{x}_2 \, S^2}
\delta^2( {\bf p}_{T\,(Z)} - {\bf p}_{T\,\rm (jet)} )\; .
\label{lotex}
\end{eqnarray}
In Eq. (\ref{lotex}) $S=(p_1+p_2)^2$ is the squared center-of-mass energy, $f(x_i,\mu)$ are the parton distribution functions, $\bar{x_1} = p^+_1/P^+_1$,  $\bar{x_2} = p^-_2/P^-_2$ are the initial-state parton momentum fractions that are fully determined at tree level, and $|M|^2$ are the relevant squared matrix elements.  The constraint on the transverse momentum of the jet is exact only at this order and only at the partonic level.

As mentioned briefly above, in order to make connection to experiment we will focus on the leptonic decay products of the $Z^0$ boson.  For this reason, one must also include the contribution of virtual photon-tagged jets in the invariant mass range around the $Z^0$ peak.  In our application to p+p and A+A collisions at the LHC we consider dilepton pairs in the invariant mass range $m_Z \pm 3\Gamma_z$, where $m_Z = 91.2$~GeV and $\Gamma_z = 2.5$ GeV~\cite{pdg}.  In this invariant mass range, the contribution from the $Z^0$ dominates that of the virtual photon by roughly two orders of magnitude. To evaluate the tagged cross section in the dilepton decay channel, we use a Monte Carlo simulation to generate the isotropic $\mu^++\mu^-$ distribution in the $Z^0/\gamma^*$ rest frame and boost this distribution back to the laboratory frame. Branching ratios are taken from Ref.~\cite{pdg}. Our approach allows us to precisely match the kinematic detector acceptance cuts. Further details of the Dalitz decay implementation are given in appendix~\ref{dalitz}.

A comparison of our LO result for jets associated with $Z^0$/$\gamma^*\rightarrow \mu^++\mu^-$ in p+${\rm \bar{p}}$ collisions at $\sqrt{s} = 1.96$~TeV with experimental measurements at the Fermi Lab Tevatron Collider \cite{Abazov:2008ez} are shown in Figure~\ref{ourdata}. The experimentally measured muon pairs are in an invariant mass range of $65$~GeV - $115$~GeV.  Additionally, the muons were required to have $p_T>15$ GeV and rapidity $|y| < 1.7$.  Jets were experimentally reconstructed using a midpoint cone algorithm with Lorentz-invariant cone size $R = \sqrt{(\Delta \phi)^2 + (\Delta y)^2} = 0.5$. Furthermore, jets were required to have $p_T > 20$ GeV and  $|y| < 2.8$. The tree level calculation under predicts the magnitude of the cross section by about $30\%$ but describes well its shape both versus transverse momentum and rapidity. The (yellow) band illustrates the sensitivity of the  $Z^0$/$\gamma^*$+jet cross section to the standard $\mu= (m_T/2,\, 2 m_T )$ variation of the renormalization and factorization scales, where  $m_T = \sqrt{m_Z^2 + p_T^2}$.  It can be taken only as a rough indicator of the size of one-loop corrections.

\begin{figure*}
\centerline{
\includegraphics[width = 0.43\linewidth]{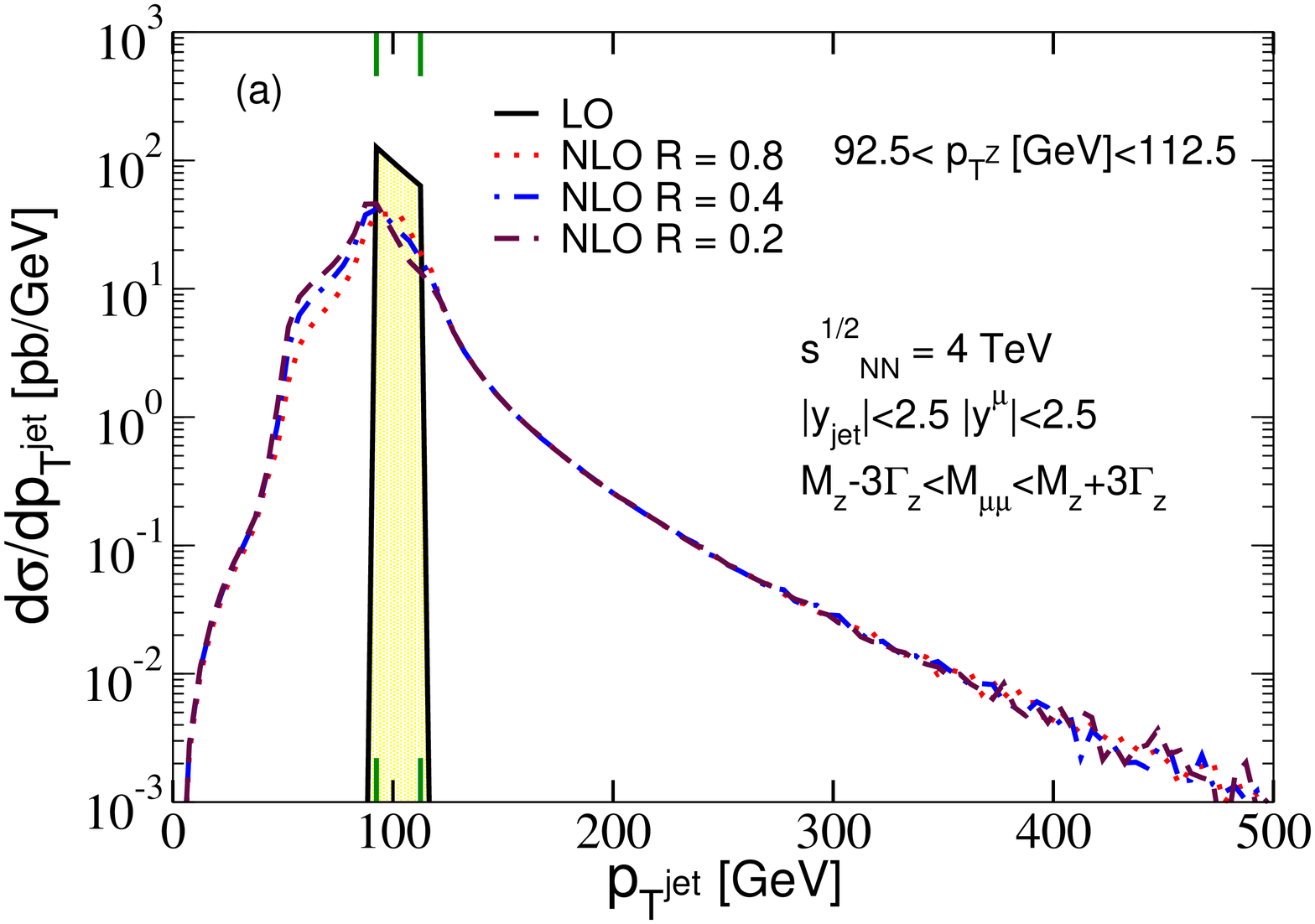}\hskip0.04\linewidth
\includegraphics[width = 0.43\linewidth]{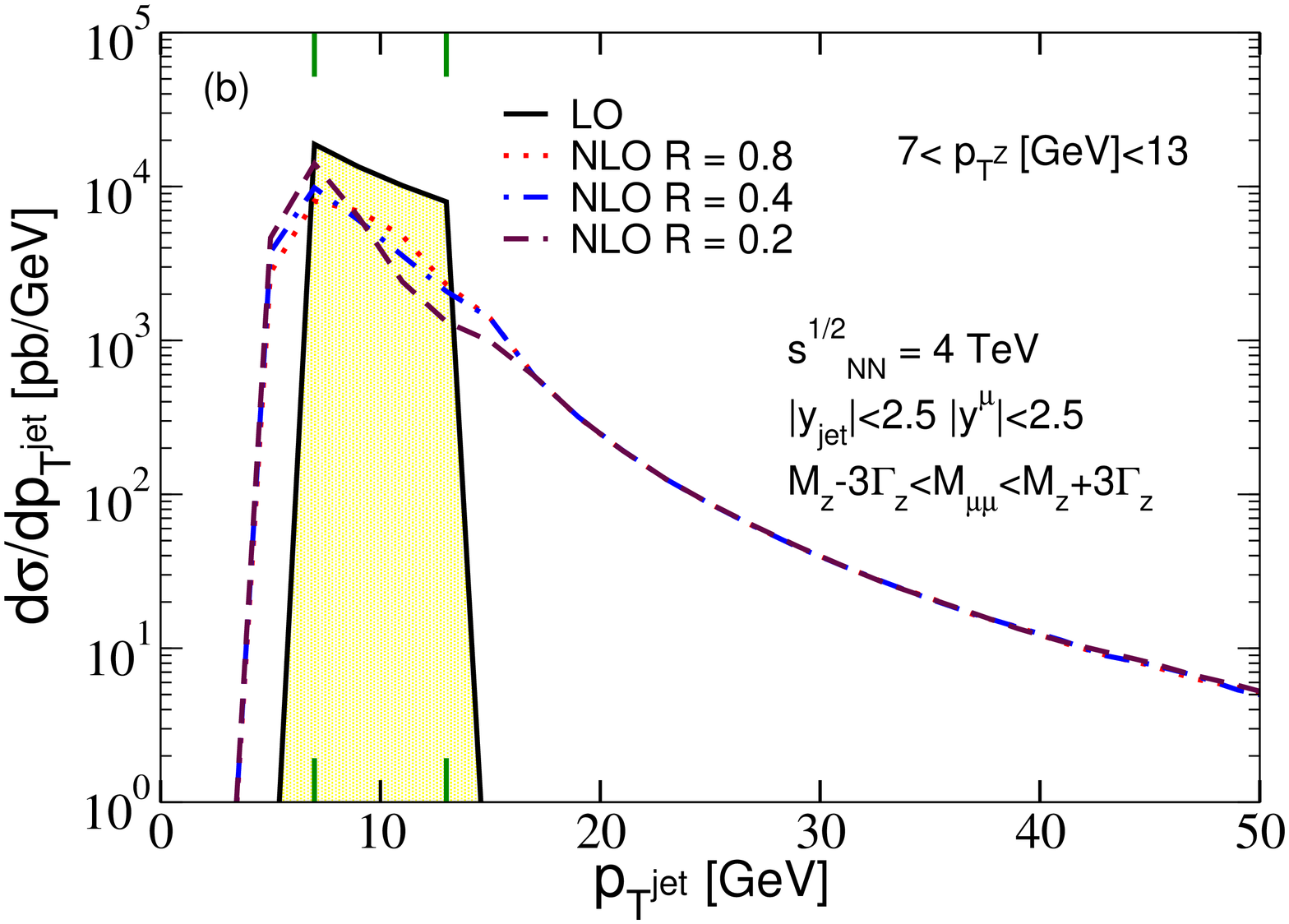}
}
\caption{(Color online) Transverse momentum distributions of a jet associated with $Z^0/\gamma^*(\rightarrow \mu^++\mu^-)$ tag
to ${\cal O}(G_F\alpha_s)$ and ${\cal O}(G_F\alpha_s^2)$. Left and right panels show results for two transverse momentum cuts $92.5\;{\rm GeV} < p_T < 112.5$~GeV  and $7\;{\rm GeV} < p_T < 13$~GeV  on the tagging particle, respectively.}
\label{shmearing}
\end{figure*}

At next-to-leading order in $\alpha_s$ the $Z^0$/$\gamma^*$+jet cross section can be written schematically as:
\begin{eqnarray}
\label{nlotex}
&& \frac{d \sigma}{d y_{(Z)} \, d y_{\rm(jet)} \, d^2 {\bf p}_{T\,(Z)} d^2 {\bf p}_{T\, \rm (jet)} }   =
\nonumber \\
&&
\frac{d^{(2)} \sigma}{d y_{(Z)} \, d y_{\rm(jet)} \, d^2 {\bf p}_{T\,(Z)} d^2 {\bf p}_{T\, \rm (jet)} }
+ \frac{1}{2!} \int dy_3 d^2{\bf p}_{T\, 3} \nonumber \\
&& \times \frac{d^{(3)} \sigma}{d y_{(Z)} \, d y_{\rm(jet)}\,
d y_{3}\, d^2 {\bf p}_{T\,(Z)} d^2 {\bf p}_{T\,\rm (jet)} d^2 {\bf p}_{T\,3} }\, .
\end{eqnarray}
The two-particle final state represents the tree-level contribution and NLO virtual corrections. In the three-particle final state one of the jets is integrated out. For the NLO cross section evaluation we make use of the publicly available Monte Carlo for FeMtobarn processes (MCFM)~\cite{Campbell:2002tg}, (available online at http://mcfm.fnal.gov/).  MCFM provides
one-loop results for many QCD processes of interest to the hadron collider physics community. Comparison between the MCFM predictions to LO and to NLO and the D0 Tevatron II data is given in Figure~\ref{tevdata}. At tree level the calculated cross sections are identical to our results, up to differences between the CTEQ and MSTW parton distribution functions~\cite{Pumplin:2002vw,Martin:2009iq}.  At NLO, the calculation is noticeably closer to the experimental data and within the systematic uncertainty of the measurement \cite{Abazov:2008ez} (not shown in Figures~\ref{ourdata} and~\ref{tevdata}).

\begin{table}[!b]
\caption{Mean $p_T$ and $\Delta \langle p_T \rangle$ for $Z^0/\gamma^*$-tagged jets at the LHC. } 
\centering 
\begin{tabular}{c c c c c c c} 
\hline\hline 
 $p_{T\, (Z)} $ [GeV] &    & LO & R = 0.2 & R = 0.4 & R = 0.8 \\ [0.5ex] 
\hline   
 7-13 &  \!\!\! $\langle p_{T\, \rm (jet)} \rangle$ [GeV] & 9.3 & 8.7 & 9.4  & 9.8\\[.5ex] 
& \!\!\!\!\!\!\! $\Delta p_{T\, \rm (jet)}$ [GeV] & 2.2 & 4.5 & 4.7 & 4.6\\[.5ex]
 92.5-112.5 &  \!\!\! $\langle p_{T\, \rm (jet)} \rangle$ [GeV] & 100.8 & 93.9 & 96.6  & 100.1\\[.5ex] 
&  \!\!\!\!\!\!\! $\Delta  p_{T\, \rm (jet)}$ [GeV] & 7.0 & 25.2 & 24.9 & 24.2\\ [1ex] 
\hline 
\end{tabular}
\label{stats} 
\end{table}

The main advantage of the next-to-leading order $Z^0/\gamma^*$+jet calculation is the ability to precisely predict the transverse momentum distribution of jets associated with a dimuon tag in a narrow $p_T$ interval. Beyond tree level, the momentum constraint that the $Z^0$ boson measurement provides is compromised by parton splitting and Z-strahlung processes.  We demonstrate this in Figure \ref{shmearing}, which shows the single differential cross section for jets tagged with $Z^0$/$\gamma^*\rightarrow \mu^+\mu^-$ in p+p collisions at $\sqrt{s_{NN}} = 4$ TeV.  Our choice of center-of-mass energy is motivated by the anticipated capabilities of the heavy-ion program at the LHC in the fall of 2010 \cite{wwnd}.  We implement acceptance cuts of $|y| < 2.5$ for both jets and final-state muons, and, as mentioned above, constrain the invariant mass of the muon pair to the interval $m_Z \pm 3\Gamma_z$, where $m_Z = 91.2$ GeV and $\Gamma_z = 2.5$ GeV.  This is the kinematic acceptance range in which we will evaluate all results that follow.

For the cross section shown in Figure \ref{shmearing}, the tagging $Z^0$/$\gamma^*$ is required to have $92.5\; {\rm GeV} < p_T < 112.5$~GeV (left panel) or $7\; {\rm GeV} < p_T < 13$~GeV (right panel).  The LO result restricts the $p_T$ of the jet to lie exactly within this interval, consistent with Eq.~(\ref{lotex}).  As seen in Figure~\ref{shmearing}, at NLO  the  deviations from this naive relation are very significant. We have included results for three different values of the jet cone radius, $R=0.2, \; 0.4, 0.8$. The variation of the cross section with $R$ around $p_{T\, \rm (jet)} \sim p_{T\,(Z)}$ arises from the  interplay between the amount of energy that is contained in the jet and the number of reconstructed jets (one or two).
For $p_{T\, \rm (jet)}\gg p_{T(Z)}$ or $p_{T\, \rm(jet)}\ll p_{T\, (Z)}$ the two final-state partons are well-separated and identified as different jets. The falloff of the differential cross section relative to its peak value at $p_{T\, \rm (jet)} = p_{T\,(Z)}$ is then controlled by the QCD splitting kernel (the part related to the large lightcone parton momentum)  and there is no dependence on the cone radius.

In order to quantify the inability of the $Z^0/\gamma^*$ tag to constrain the momentum of the jet we calculate the mean $p_T \equiv \langle p_{T\, \rm(jet)}\rangle$ and standard deviation
$\Delta p_{T\, \rm(jet)}  = \sqrt{\langle p_{T\, \rm(jet)}^2 \rangle - \langle p_{T\, \rm(jet)} \rangle^2}$ for each of the curves in Figure~\ref{shmearing}.  The results are presented in Table~\ref{stats}. The standard deviation for the LO curves is not strictly  zero because of the finite $p_T$ width of the tagging $Z^0$/$\gamma^*$ bin.  The NLO curves exhibit a similar $\langle p_{T\, \rm(jet)}\rangle$ as the LO result, with  $\langle p_{T\, \rm(jet)} \rangle$
increasing as the cone radius increases. However, in going from LO to NLO there is a significant jump
in $\Delta p_{T\, \rm(jet)}$. The width of the jet momentum distribution  quadruples for the more energetic tag. The very large values of
$ \Delta p_{T\, \rm(jet)} / \langle p_{T\, \rm(jet)} \rangle   \sim 25 \%$ at NLO create serious complications for experimentally tagging the initial associated jet energy in both p+p and A+A collisions. While additional cuts can be considered, such as the requirement for a single jet within the
experimental acceptance that is exactly opposite the tagging particle in azimuth, these will reduce the already small projected multiplicity for this final state in heavy ion reactions.

\section{Medium-Induced parton showers from final-state interactions in A+A reactions}\label{eaa}

\subsection{Radiative energy loss of fast partons in the QGP}

The principle difference between jet physics in proton-proton collisions and
jet-physics in nucleus-nucleus collisions is the contribution of final-state
inelastic quark and gluon interactions in the QGP to the formation of parton
showers \cite{Vitev:2008rz,Vitev:2009rd,Vitev:2008bx,Neufeld:2010sz}. In this work we use
a theoretical approach developed to address the problem of parton energy loss
at RHIC and at the LHC \cite{Gyulassy:2000fs,Vitev:2007ve}.

To evaluate the medium-induced bremsstrahlung, it is important to keep track
of the evolution of the gluon transverse momentum ${\bf k}$ in a plane perpendicular
to the direction of jet propagation. Such a ${\bf k}$ may arise from a single hard scattering or from
multiple soft scatterings. The acceleration of the color charges in the 2D transverse
plane generates color currents whose detailed interference pattern determines the
strength of the non-Abelian Landau-Pomeranchuk-Migdal effect~\cite{Gyulassy:2003mc,Baier:1996sk,Zakharov:1997uu,Gyulassy:2000fs,Guo:2000nz}.
Let us denote by:
\beqar
\label{hprop}
{\bf H}&=&{{\bf k} \over {\bf k}^2 }\; , \qquad \qquad  \\[1.ex]
{\bf C}_{(i_1 \cdots i_m)}&=&{{\bf k} - {\bf q}_{i_1} -
\cdots -{\bf q}_{i_m}
\over ({\bf k} - {\bf q}_{i_1} -
\cdots -{\bf q}_{i_m}   )^2 } \;,\qquad
 \\[1.ex]
{\bf B}_{i_1} &= &{\bf H} - {\bf C}_{i_1} \; , \qquad \\[1.ex]
{\bf B}_{(i_1  \cdots i_m )(j_1j_2 \cdots i_n)} &=&
{\bf C}_{(i_1 \cdots j_m)} - {\bf C}_{(j_1 j_2 \cdots j_n)} \;, \;\;
\eeqar{hbgcdef}
the Hard, Cascade, and Bertsch-Gunion propagators in the transverse momentum space~\cite{Gyulassy:2000fs}.
In Eqs. (\ref{hprop})-(\ref{hbgcdef})  ${\bf q}_{i}$ are the momentum transfers from the medium.
Another important quantity, which enters the bremsstrahlung spectrum, is the formation time of the gluon,
$\tau_f$, at the radiation vertex. When compared to the separation between the scattering centers
$\Delta z_j  = z_j - z_{j-1}$, which can fluctuate from zero up to the size of the medium $L$,  it determines
the degree of coherence present in the multiple scattering process. We introduce the following
notation~\cite{Gyulassy:2000fs}:
\beqar
\tau_0^{-1} = \omega_0 &=& \frac{ {\bf k}^2 }{k^+}\; , \qquad
\qquad  \\[1.ex]
\tau_{i_1 }^{-1} =  \omega_{i_1}
&=&  \frac{ ({\bf k} - {\bf q}_{i_1} )^2  }{k^+} \;, \qquad  \\[1.ex]
\tau_{(i_1  \cdots i_m)}^{-1} =  \omega_{(i_1 \cdots i_m)}
&=&  \frac{ ({\bf k} - {\bf q}_{i_1} -
 \cdots -{\bf q}_{i_m} )^2  }{k^+} \;, \qquad
\eeqar{phases}
where $k^+$ is the gluon's large lightcone momentum.

For final-state (FS) interactions, the double differential distribution of medium-induced gluons
reads \cite{Vitev:2007ve}:
\begin{widetext}
\beqar
k^+ \frac{dN^g(FS)}{dk^+ d^2 {\bf k} } &=& \frac{C_R \alpha_s}{\pi^2}
\sum_{n=1}^{\infty}  \left[ \prod_{i = 1}^n  \int
\frac{d \Delta z_i}{\lambda_g(z_i)}  \right]
\left[ \prod_{j=1}^n \int d^2 {\bf q}_j \left( \frac{1}{\sigma_{el}(z_j)}
\frac{d \sigma_{el}(z_j) }{d^2 {\bf q}_j}
-  \delta^2 ({\bf q}_j) \right)    \right] \nonumber \\
&& \times \;  \left[ -2\,{\bf C}_{(1, \cdots ,n)} \cdot
\sum_{m=1}^n {\bf B}_{(m+1, \cdots ,n)(m, \cdots, n)}
\left( \cos \left (
\, \sum_{k=2}^m \omega_{(k,\cdots,n)} \Delta z_k \right)
-   \cos \left (\, \sum_{k=1}^m \omega_{(k,\cdots,n)}
\Delta z_k \right) \right)\; \right]  \;.   \qquad
\eeqar{full-final}
\end{widetext}
In Eq. (\ref{full-final}) $\sum_2^1 \equiv 0$ and  ${\bf B}_{(n+1, n)} \equiv {\bf B}_n$ is understood.
 In the case of final-state interactions, $z_0 \approx 0$ is the point of the initial hard scattering and
$z_L = L$ is the extent of the medium.  The path ordering of the interaction points,
$z_L > z_{j+1} > z_j > z_0$, leads to the constraint $\sum_{i=1}^n \Delta z_i  \leq  z_L $.
One implementation of this condition would be $\Delta z_i \in [\, 0,z_L -\sum_{j=1}^{i-1} \Delta z_j \, ]$
and it is implicit in Eq.~(\ref{full-final}). We also note that  $C_R = C_F\equiv 4/3$ for quark jets
and $C_R = C_A\equiv 3$ for gluon jets. The transverse momentum transfers from the medium are averaged over the
normalized scattering cross section   $[{1}/{\sigma_{el}(z_j)}]
{d \sigma_{el}(z_j) }/{d^2 {\bf q}_j} $.

\subsection{Numerical methods and QGP properties}
\label{subsec:QGP}

Results relevant to the LHC phenomenology  are calculated using full numerical evaluation  of
the medium-induced contribution to the parton showers. These affect the observed jet shapes and
the in-medium jet cross sections. Energetic inclusive jet production and tagged jet production are rare  processes
that follow binary collision  scaling $ \sim d^2N_{\rm bin}/d^2{\bf x} $. In contrast, the medium
is distributed according to the number of participants density $ \sim d^2N_{\rm part}/d^2{\bf x}$.
Soft particles that carry practically all of the energy deposited in the fireball of a heavy ion
collision cannot deviate a much from such scaling.  We take into account the longitudinal Bjorken
expansion of the QGP and the medium density as a function of proper time reads:
\begin{equation}
\rho({\bf x},\tau) = \rho({\bf x},\tau_0) \frac{\tau_0}{\tau} \;.
\label{bg}
\end{equation}
In Eq.~(\ref{bg}) the initial density at time $\tau_0$ can be related to the experimentally
measured charged particle rapidity density and the participant density in a plane perpendicular to the collision axis:
\begin{eqnarray}
\rho({\bf x},\tau_0) & = & \frac{1}{\tau_0}  \frac{d^2 ( dN^g/dy)}{d^2 {\bf x} }
\approx  \frac{1}{ \tau_0 } \frac{3}{2}
\left| \frac{d\eta}{dy} \right| \frac{d^2 (dN^{ch}/d\eta) }{d^2 {\bf
x}}  \nonumber \\
&=& \kappa  \frac{d^2 N_{\rm part}}{d^2 {\bf x} }  \;  .
\label{ydep}
\end{eqnarray}

\begin{figure*}[!t]
\centerline{
\includegraphics[width = 0.40\linewidth]{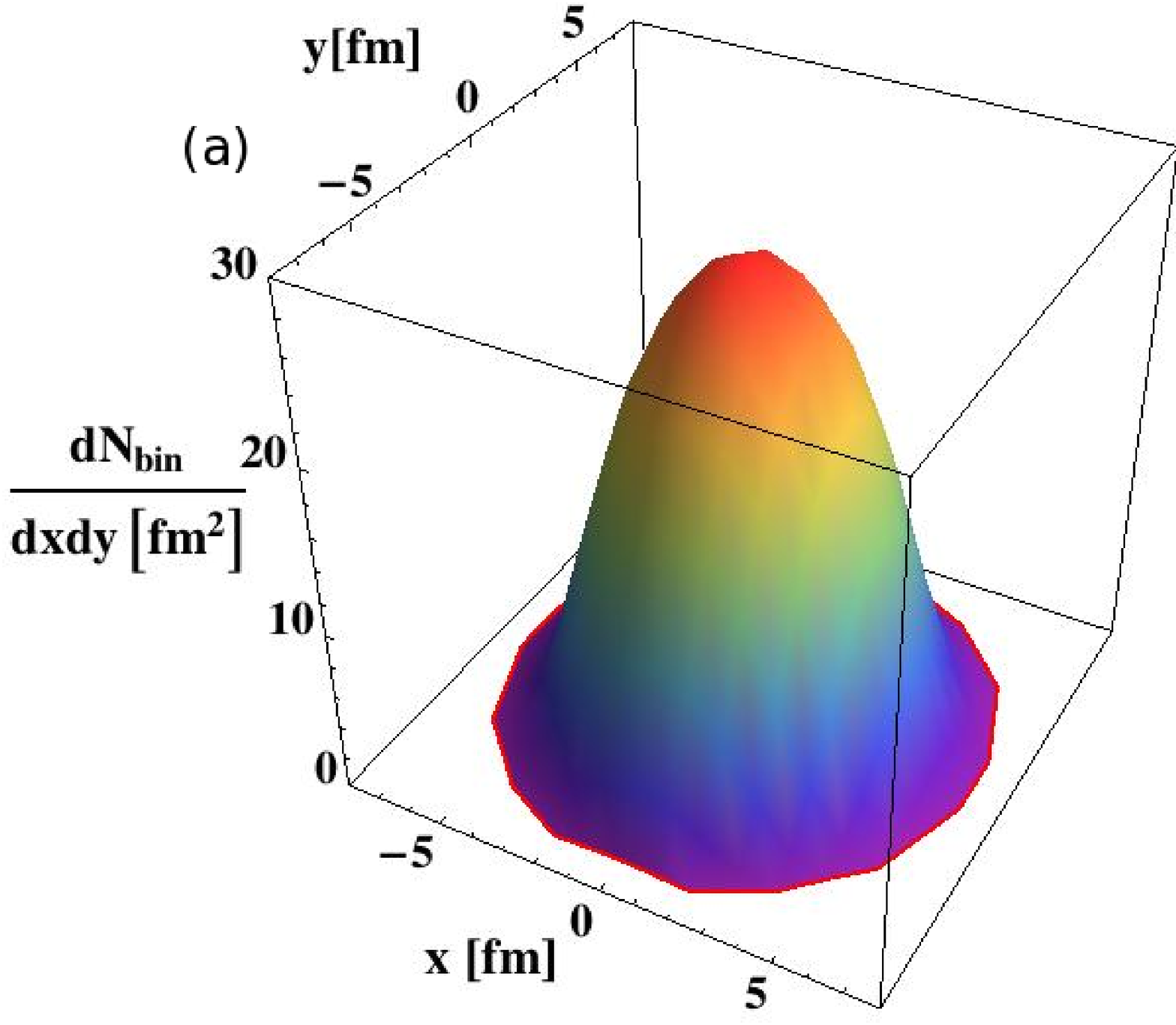}\hskip0.05 \linewidth
\includegraphics[width = 0.45\linewidth]{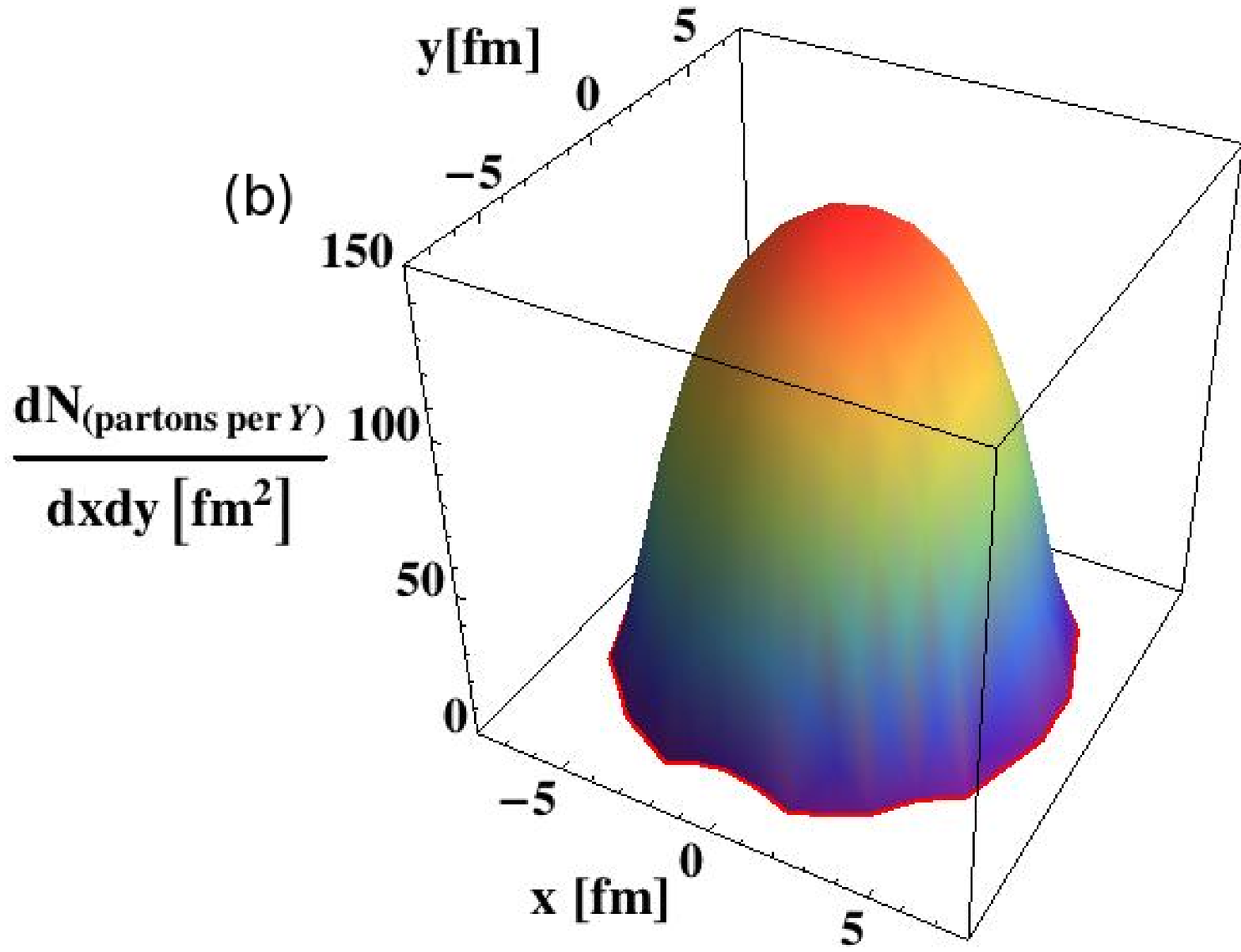} }
\caption{(Color online) The left panel shows the binary collision density in central Pb+Pb collisions
at the LHC from an optical Glauber model with $\sigma_{in} =65$~mb. The right panel shows the transverse
density of soft partons/particles, which is proportional to the participant density.}
\label{partbin}
\end{figure*}

As an illustration, we show in the left panel of Figure~\ref{partbin} the distribution of jets
in central ($b=3$~fm) Pb+Pb collisions at the LHC. This binary collision density is evaluated
using an optical Glauber model with an inelastic scattering cross section $\sigma_{in} =65$~mb.
The density of QGP partons/soft particles in the plane transverse to the collision axis is
shown in the right panel of Figure~\ref{partbin}. We note
that for the constituents of the medium we use parton/hadron duality, see Eq.~(\ref{ydep}). Specifically,
$dN^g/dy =  dN^{ch+neut.}/dy = 2800$. Assuming local thermal equilibrium, for a gluon-dominated plasma
we find:
\begin{equation}
T(\tau, {\bf x}) = \  ^3\!\sqrt{ \pi^2 \rho(\tau,{\bf
x}) / 16 \zeta(3) }\;, \tau > \tau_0 \; .
\label{tempdet}
\end{equation}
The Debye screening scale is then given by $m_D^2 = 4\pi \alpha_sT^2$ and
the relevant gluon mean free path is easily evaluated:  $\lambda_g = 1/ \sigma^{gg} \rho$ with
$\sigma^{gg} = (9/2) \pi \alpha_s^2 / m_D^2$.

In our simulation we generate in-plane jets, $ \phi_{\rm jet} - \phi_{\rm reaction\; plane} = 0 $, since
in central Pb+Pb collisions the interaction region is nearly azimuthally symmetric. We evaluate the differential
bremsstrahlung spectrum  Eq.~(\ref{full-final}) to the lowest non-trivial interference  between the
gluon emission from the hard jet production and the subsequent interactions in the
QGP~\cite{Gyulassy:2000fs}. In our simulation the allowed gluon phase space is
$\Lambda_{QCD} < \omega < E$,
$\Lambda_{QCD} < k_\perp < 2 \omega $~\footnote{This condition allows for the deflection of the jet and
can be also derived from the finite rapidity range constraint $ 0 < y_g < y_{\rm jet} $ for the emitted gluon},
and the transverse momentum transfers between the jet and the medium are in the
interval $0 < q_{\perp\, i} < \sqrt{s/4} = \sqrt{m_D E/2 }$~\cite{Vitev:2007ve}. Finally, we note that
we study two coupling strengths between the fast partons and the medium, $\alpha_s = 0.3,\, 0.5$,
and use a running $\alpha_s(k_T)$ for the emission vertex.

\section{Predictions for Heavy-Ion Collisions at LHC Energies}\label{results}

\begin{figure}
\vskip0.01\linewidth
\centerline{
\includegraphics[width=0.45\textwidth]{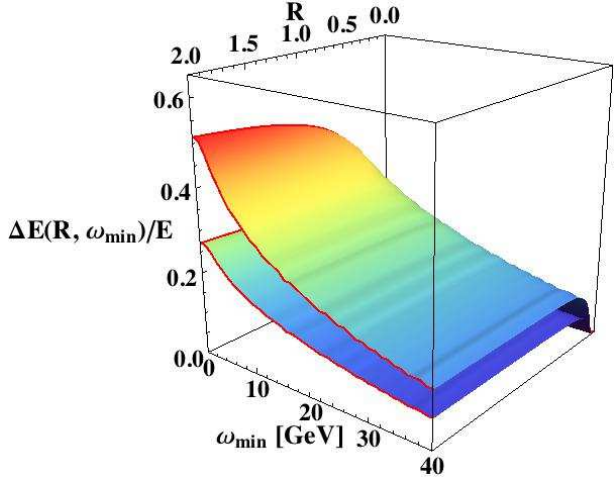} }
\caption{(Color online) Fractional energy $\Delta E / E$  lost by 100~GeV quark and gluon jets inside a cone of radius $R$ via
gluon emission of $\omega > \omega_{\rm min}$ as a result of final-state interactions  in the QGP created
in central Pb+Pb reactions at the LHC.}
\label{deexample}
\end{figure}

With the numerically expensive simulations of energetic quark and gluon propagation through the expanding QGP medium created in central Pb+Pb reactions at the LHC
completed, we proceed to evaluate the $Z^0/\gamma^*$+jet cross section in these ultrarelativistic nuclear collisions.  In our approach, partonic energy loss is incorporated through two crucial ingredients.  The first is the fraction of lost energy that falls within a certain cone radius $R$ and is carried away by gluons of energy above a minimum  energy cut-off
$\omega_{\rm min}$ \cite{Vitev:2008rz}.  We denote this fraction by:
\eqf{\label{fdeaf}
f_{q,g} = f_{q,g}(R,\omega_{\min},E) = \frac{\Delta E(R,\omega_{\min},E)}{\Delta E(R^\infty,0,E)} \, .
}
In Eq.~(\ref{fdeaf})  $E$ is the initial energy of the  jet parent  parton, and the subscript $q,g$ indicates whether this parton is a quark or a gluon.
The function $f_{q,g}$ describes the fraction of partonic energy loss that will be redistributed inside the jet for fixed jet reconstruction parameters
$R$ and $\omega_{\rm min}$. $\Delta E(R,\omega_{\min},E)$ is obtained from the double differential spectrum of medium-induced bremsstrahlung
${d I^g}/{d\omega d r}$ as follows:
\eqf{
\Delta E(R,\omega_{\min},E) = \int_{\omega_{\min}}^E d \omega \int_0^R  d r \frac{d I^g}{d\omega d r}(\omega,r)\, .
}
Figure \ref{deexample} shows an example of the mean  ${\Delta E(R,\omega_{\min},E)/E}$ for 100 GeV quark and gluon jets that were created
in central Pb+Pb reactions at the LHC and have propagated through the QGP medium.

The second ingredient we will need is $P_{q,g}(\epsilon)$, which is the probability distribution for the fractional
energy loss of an energetic parton due to multiple gluon emission, $\epsilon = \sum_i \omega_i/E$. The evaluation of
this distribution from the calculated medium-induced bremsstrahlung spectrum is described elsewhere \cite{Vitev:2005he}.
Knowledge of $P_{q,g}(\epsilon)$ is essential for the probabilistic superposition of vacuum and medium-induced parton
showers and the calculation of inclusive and tagged jet cross sections in heavy in reactions.

\subsection{Evaluating the suppressed $Z^0/\gamma^*$+jet cross section}

Consider the invariant cross section for $Z$/$\gamma^*$-tagged jets in p+p collisions at an arbitrary order in perturbation theory, which we write as:
\eqf{
\frac{d\sigma}{d^2\bald{p}_{(Z)}d^2\bald{p}_{\rm (jet)}} =
\int_{y_{(Z)\min}}^{y_{(Z)\max}} dy_{(Z)}
\int_{y_{\rm(jet)\min}}^{y_{\rm(jet)\max}}  dy_{\rm(jet)} \\
\frac{d\sigma}{dy_{(Z)}d^2\bald{p}_{(Z)}dy_{\rm(jet)} d^2\bald{p}_{\rm(jet)}}.
}
Here, we are interested in obtaining the medium-modified jet cross section per binary $N+N$ scattering in terms of the above expression.
In particular, medium-induced energy loss does not affect on average the direction of the jet (and $y_{\rm(jet)}$) and its effects are contained primarily in $\bald{p}_{\rm(jet)}$. We define $\bald{p}_{Q} = \bald{p}_{\rm(jet)} (1 -  ( 1 - f_{q,g} ) \epsilon)$
which is the reduced momentum accounting for medium-induced energy loss that falls outside a certain cone radius and energy cut-off.  In the limit of zero energy loss ($\epsilon \rightarrow 0$) or very large jet cone radius ($f_{q,g} \rightarrow 1$)  one recovers $\bald{p}_{Q} \rightarrow \bald{p}_{\rm(jet)}$.  Accounting for the fact that quark and gluon fractional energy loss is described by a probability distribution $P_{q,g}(\epsilon)$ and the transverse momentum Jacobian: $$\left| \frac{d^2{\bf p}_{\rm(jet)}}{d^2{\bf p}_{(Q)}}  \right| = \frac{1}{[1 - (1-f_{q,g}(\omega_{\min},R))\epsilon]^2}\;,$$ for the quenched jet cross section we find:
\begin{eqnarray}
\frac{d\sigma}{d^2 \bald{p}_{(Z)} d^2 \bald{p}_Q} \!\!\!&=& \!\!\!
\sum_{q,g} \int d \epsilon \; P_{q,g}(\epsilon) \frac{1}{[1 - (1-f_{q,g}(\omega_{\min},R))\epsilon]^2} \nonumber \\
&& \times \frac{d\sigma^{q,g}}{d^2 \bald{p}_{(Z)} d^2 \bald{p}_{\rm(jet)}}  \left(\frac{\bald{p}_{Q}}
{[1 - (1-f_{q,g})\epsilon)]}\right) \;. \qquad
\label{2Dquench}
\end{eqnarray}
The physical meaning of Eq.~(\ref{2Dquench}) is that the observed tagged jet cross section in A+A reactions is a probabilistic superposition of
cross sections for jets of higher initial transverse energy. This excess energy is then redistributed outside of the jet due to strong
final-state interactions.
Here, $P_{q,g}(\epsilon)$ and $f_{q,g}(\omega_{\min},R)$ are obtained using the GLV formalism for evaluating the medium-induced gluon bremsstrahlung~\cite{Gyulassy:2000fs,Vitev:2007ve}. Also, $d\sigma^{q,g}/{d^2 \bald{p}_{(Z)} d^2 \bald{p}_{\rm(jet)} }$ are the differential cross sections for away-side quark and gluon jets, respectively. We calculate the relative fraction of these jets to lowest order in perturbation theory.  Even though at next-to-leading order one might expect a small correction to these results, we show in Appendix~\ref{qgvariation} that a variation of the ${\cal O}(\alpha_s)$ in the relative quark to gluon jet fraction does not affect the predicted quenching of $Z^0/\gamma^*$-tagged jets.

It is often the case that the $Z^0/\gamma^*$+jet final-state channel is measured without placing restrictions on the momentum of the vector boson~\cite{Abazov:2008ez}. In this case, integrating over $\bald{p}_{(Z)}$ in the LHS and RHS of Eq.~(\ref{2Dquench}) we obtain:
\begin{eqnarray}
\frac{d\sigma}{ d^2 \bald{p}_Q} \!\!\!&=& \!\!\!
\sum_{q,g} \int d \epsilon \; P_{q,g}(\epsilon) \frac{1}{[1 - (1-f_{q,g}(\omega_{\min},R))\epsilon]^2} \nonumber \\
&& \times \frac{d\sigma^{q,g}}{ d^2 \bald{p}_{\rm(jet)}}  \left(\frac{\bald{p}_{Q}}
{[1 - (1-f_{q,g})\epsilon)]}\right) \;. \qquad
\label{2DquenchIncl}
\end{eqnarray}
For all practical purposes, Eq.~(\ref{2DquenchIncl}) contains the same physics as the suppression of the cross section in inclusive jet measurements~\cite{Vitev:2008rz,Vitev:2009rd,Vitev:2008bx}.
We will demonstrate this shortly in our numerical results section.

In the special case of a tree level calculation for the elementary quark and gluon jet tagged cross sections we recognize that:
$$ \frac{d\sigma}{d^2\bald{p}_{(Z)} d^2\bald{p}_{\rm(jet)}}
 = \frac{d\sigma}{d^2\bald{p}_{(Z)}} \delta^2 ({\bf p}_{(Z)}-{\bf p}_{\rm (jet)})\;.$$
Substituting this result in Eq.~(\ref{2Dquench}), after straightforward algebraic manipulation we find:
\begin{eqnarray}
\frac{d\sigma}{ d^2 \bald{p}_{(Z)} d^2 \bald{p}_Q }  \!\!&=& \!\! \sum_{q,g}
\frac{\delta(\phi_{(z)}-\phi_{\rm(jet)}-\pi)}{p_{T\, (Q)} p_{T\, (Z)} (1-f_{q,g})}
\frac{d\sigma^{q,g}}{d^2\bald{p}_{(Z)}} \nonumber \\
&& \times \, P_{q,g}\left( \frac{1-p_{T\,(Q)}/p_{T\,(Z)}}{1-f_{q,g}} \right)  \;. \qquad
\label{2DquenchLO}
\end{eqnarray}
Note that in Eq.~(\ref{2DquenchLO}) the remaining $\delta$-function simply reflects the fact that at tree level the $Z^0/\gamma^*$ and the jet are exactly back-to-back and the angular distributions can be trivially integrated over. From the  $0\leq f_{q,g}(\omega_{\min},R)\leq 1$ and the properties of $P_{q,g}(\epsilon)$ one finds that the double differential cross section
exists in the region $f_{q,g} p_{T\, (Z)} \leq p_{T\,(Q) }\leq p_{T\, (Z)}$.

It is instructive to investigate the $f_{q,g}\rightarrow 1$ limit in Eq.~(\ref{2DquenchLO}). For $p_{T\,(Q)} < p_{T\,(Z)}$, $P_{q,g}(\epsilon >1) \equiv 0$ ensures a lack of singularity. If $p_{T\, (Q)} \rightarrow p_{T\, (Z)}$ we have $\lim_{\epsilon \rightarrow 0} P_{q,g}(\epsilon) \rightarrow \exp(-\langle N^g \rangle_{q,g}) \delta(\epsilon)$~\cite{Gyulassy:2001nm}. Here, $\langle N^g \rangle_{q,g} $  is the average number of medium-induced gluons for a quark or gluon jet respectively. To interpret the resulting cross section we integrate over ${\bf p}_{T\,(Q)}$ to find:
\begin{eqnarray}
\frac{d\sigma}{d^2 \bald{p}_{(Z)}}\Bigg|_{p_{T\,(Q)}\equiv p_{T\,(Z)}}
 \!\!&=& \!\! \sum_{q,g} e^{-\langle N^g \rangle_{q,g} } \frac{d\sigma^{q,g}}{d^2 \bald{p}_{(Z)}}  \;.
\label{1DquenchLO}
\end{eqnarray}
The physical meaning of Eq.~(\ref{1DquenchLO}) is that at lowest order the coincident $Z^0/\gamma^*$+jet production with ${\bf p}_{T\,(Q)}\equiv - {\bf p}_{T\,(Z)}$ is directly proportional to the probability {\em not} to lose energy via medium-induced gluon bremsstrahlung. Eqs.~(\ref{2DquenchLO}) and (\ref{1DquenchLO}) imply that one might gain easy access to $\langle N^g \rangle_{q,g}$ and $P_{q,g}(\epsilon)$. However, higher order corrections significantly alter these relations. To disentangle these quantities/distributions from tagged jet measurements one still needs accurate baseline calculations of the selected final-state channel at NLO and, ideally, precise experimental measurements.

We conclude this section by restating in a compact form the basic results for the evaluation of inclusive and tagged jets 
in the ambiance of strongly-interacting matter:
\begin{eqnarray}
\frac{d\sigma}{d p_{T\,(Q)}} &=& \sum_{q,g}  \int_0^1 d \epsilon
\frac{P_{q,g}(\epsilon)}{[1 - (1-f_{q,g})\epsilon]} \nonumber \\
&& \times \frac{d\sigma^{q,g}}{d p_{T\,\rm(jet)}}\; , \\
\frac{d\sigma}{d p_{T\,(Z)}d p_{T\,(Q)}} &=& \sum_{q,g}  \int_0^1 d \epsilon
\frac{P_{q,g}(\epsilon)}{[1 - (1-f_{q,g})\epsilon]} \nonumber \\
&& \times\frac{d\sigma^{q,g}}{dp_{T\,(Z)}d p_{T\,\rm(jet)}} \;.
\end{eqnarray}

\begin{figure}
\vskip0.01\linewidth
\centerline{
\includegraphics[width=0.45\textwidth]{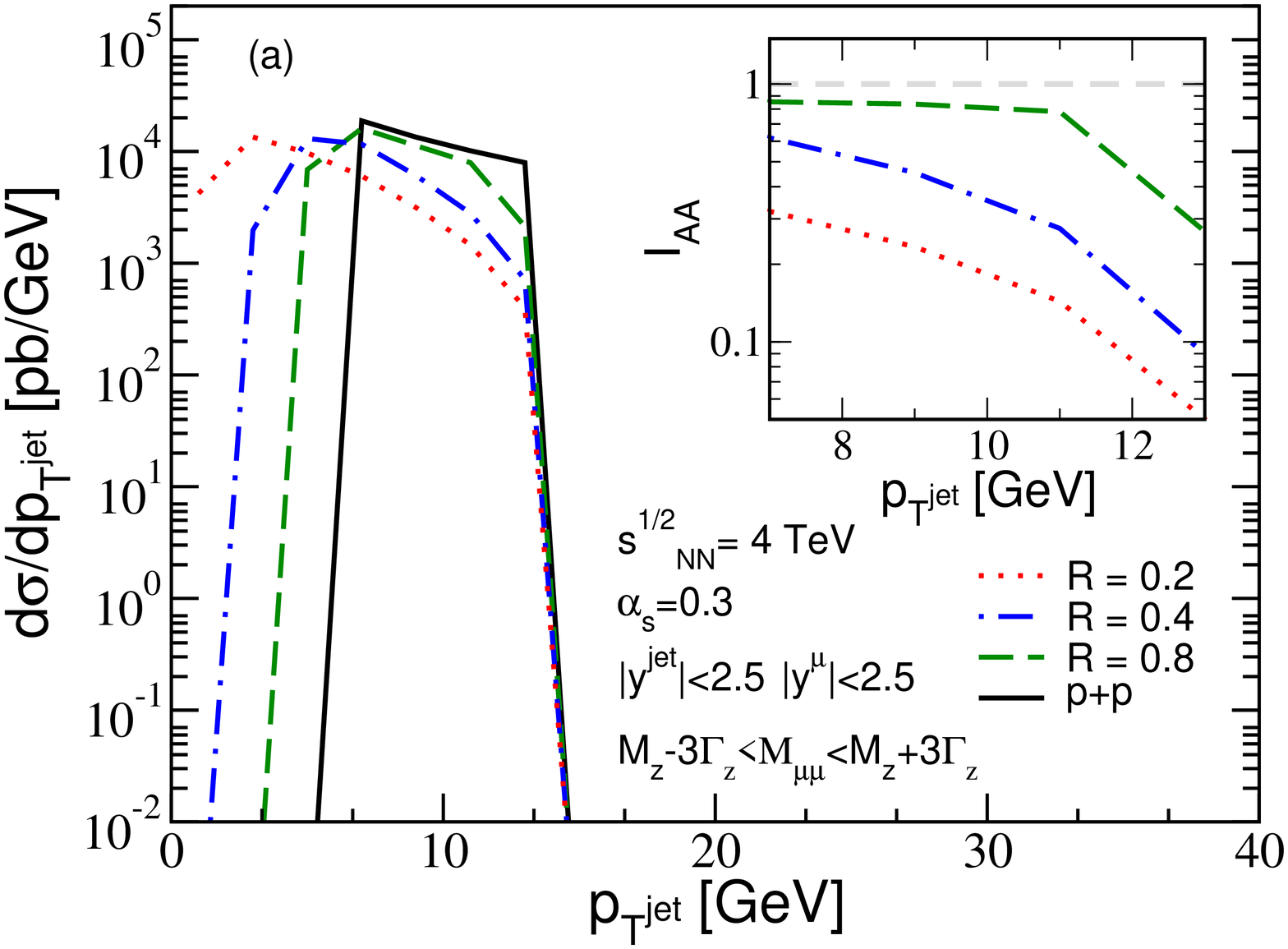}}
\vskip 0.1\linewidth
\centerline{
\includegraphics[width=0.45\textwidth]{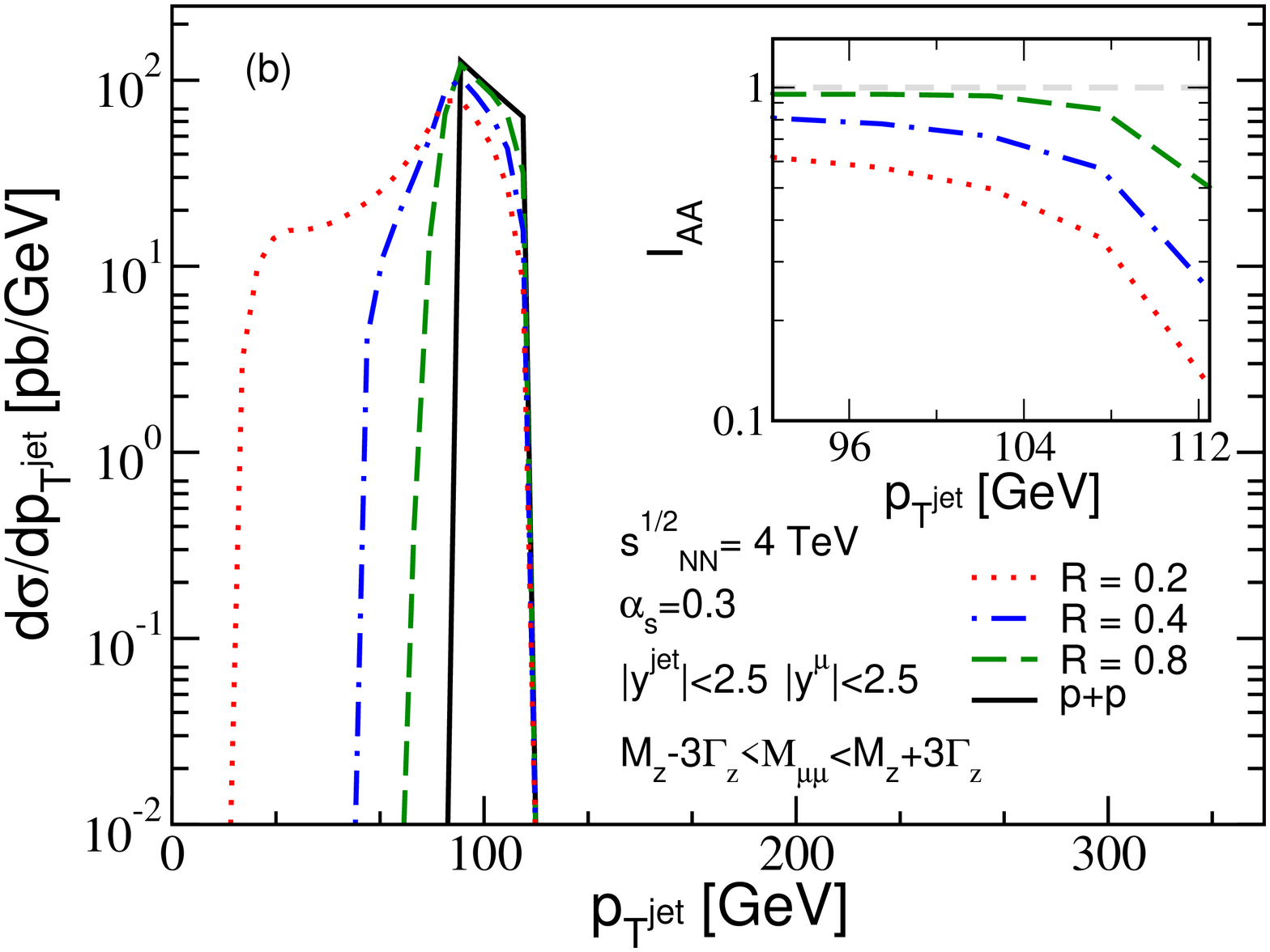} }
\caption{(Color online)  LO results for the-$p_T$ differential $Z$/$\gamma^*$+jet  cross section where the tagging $Z$/$\gamma^*$ is
required to have $7\; {\rm GeV}  < p_T < 13\; {\rm GeV}$  (top panel) and $92.5 \; {\rm GeV} < p_T < 112.5\; {\rm GeV}$ (bottom panel).
We present simulations for $\sqrt{s_{NN}}=4$~TeV p+p and central Pb+Pb collisions at the LHC for three different jet cone radii $R$. }
\label{lo_meander}
\end{figure}

\subsection{Numerical Results}\label{numres}

We now present numerical results for the QGP-modified $Z^0/\gamma^*$+jet production cross sections in central Pb+Pb collisions at the LHC at $\sqrt{s_{NN}}=4$~TeV.
As for p+p reactions, we consider jets tagged with $Z$/$\gamma^*\rightarrow \mu^++\mu^-$ using acceptance cuts of $|y| < 2.5$ for both jets and final-state muons.
Also, the invariant mass of the muon pair is required to be in the interval $m_Z \pm 3\Gamma_z$, where $m_Z = 91.2$ GeV and $\Gamma_z = 2.5$ GeV.
In this paper we focus on the effect of final-state interactions, motivated by the fact that both the large virtuality $Q^2=m_{T\, (z)}^2$ and the
range of $ m_Z < m_{T\, (z)} \ll \sqrt{s_{NN}}/2$ (at midrapidity) are expected to reduce the significance of initial-state effects relative to RHIC
kinematics~\cite{Vitev:2009rd}. We defer detailed studies of initial-state interactions to future work.

We start with the tree level results for the $p_T$-differential jet cross section where the tagging $Z$/$\gamma^*$ is required to have
$7 \; {\rm GeV} < p_T < 13\; {\rm GeV} $ and $92.5 \; {\rm GeV} < p_T < 112.5 \; {\rm GeV} $. These cross sections are shown in the top and bottom panels of
Figure \ref{lo_meander}, respectively. In the collinear factorization approach at leading order the momentum of the $Z^0/\gamma^*$ is exactly balanced by the momentum of
the jet. Furthermore, to this order the jet is represented by the energetic parent parton and the cross section is insensitive to the choice of the jet cone
radius $R$~\footnote{We defer the discussion of hadronization and ``splash out'' effects to future studies.}. In Pb+Pb reactions the redistribution of the jet energy is determined by the medium modification to the parton shower but the experimentally observable effect is controlled by the jet cone radius $R$.
At tree level, the choice of radius (and/or $\omega_{\min}$) also determines the $p_T$ range where the tagged jet cross section does not vanish,
see Eq.~(\ref{2DquenchLO}). We note that, unless stated otherwise, in our numerical simulations we assume no momentum cut on the soft particles/partons, $\omega_{\min} =0$~GeV. The distribution of jets in central Pb+Pb collisions in Figure~\ref{lo_meander} reflects the probability distribution for the fractional energy loss of partons
$\langle P(\epsilon) \rangle$ \cite{Neufeld:2010sz}, the average being over the quark and gluon jets.

The inserts in Figure \ref{lo_meander} show a measure of nuclear effects for tagged jet cross sections in heavy ion reactions relative to the
binary collision $\langle N_{\rm bin} \rangle$ scaled cross section in proton-proton collisions:
\begin{equation}
I_{AA}^{\rm jet}(R,\omega_{\min}) = \frac{1}{\langle N_{\rm bin} \rangle}\frac{d\sigma_{AA}}{d p_{T\,(Z)}d p_{T\,(Q)}}
\Bigg / \frac{d\sigma_{pp}}{d p_{T\,(Z)}d p_{T\,(\rm jet)}} \; .
\label{iaa}
\end{equation}
As we will see shortly, $I_{AA}^{\rm jet}$ is much more useful at NLO. At tree level it is limited to the $Z^0/\gamma^*$ trigger momentum range. The large
suppression toward the upper edge of transverse momentum interval is reminiscent of the $\exp(-\langle N^g \rangle_{q,g})$ factor in Eq.~(\ref{1DquenchLO})
and indicates that parton energy loss in the QGP proceeds through the emission of several semi-hard (of energies of the order of few~GeV) gluons.
Unfortunately, in realistic measurements the finiteness of the $p_T$ trigger bin itself will prevent a more accurate determination of $ \langle N^g \rangle_{q,g}$.

\begin{figure}
\centerline{
\includegraphics[width=0.45\textwidth]{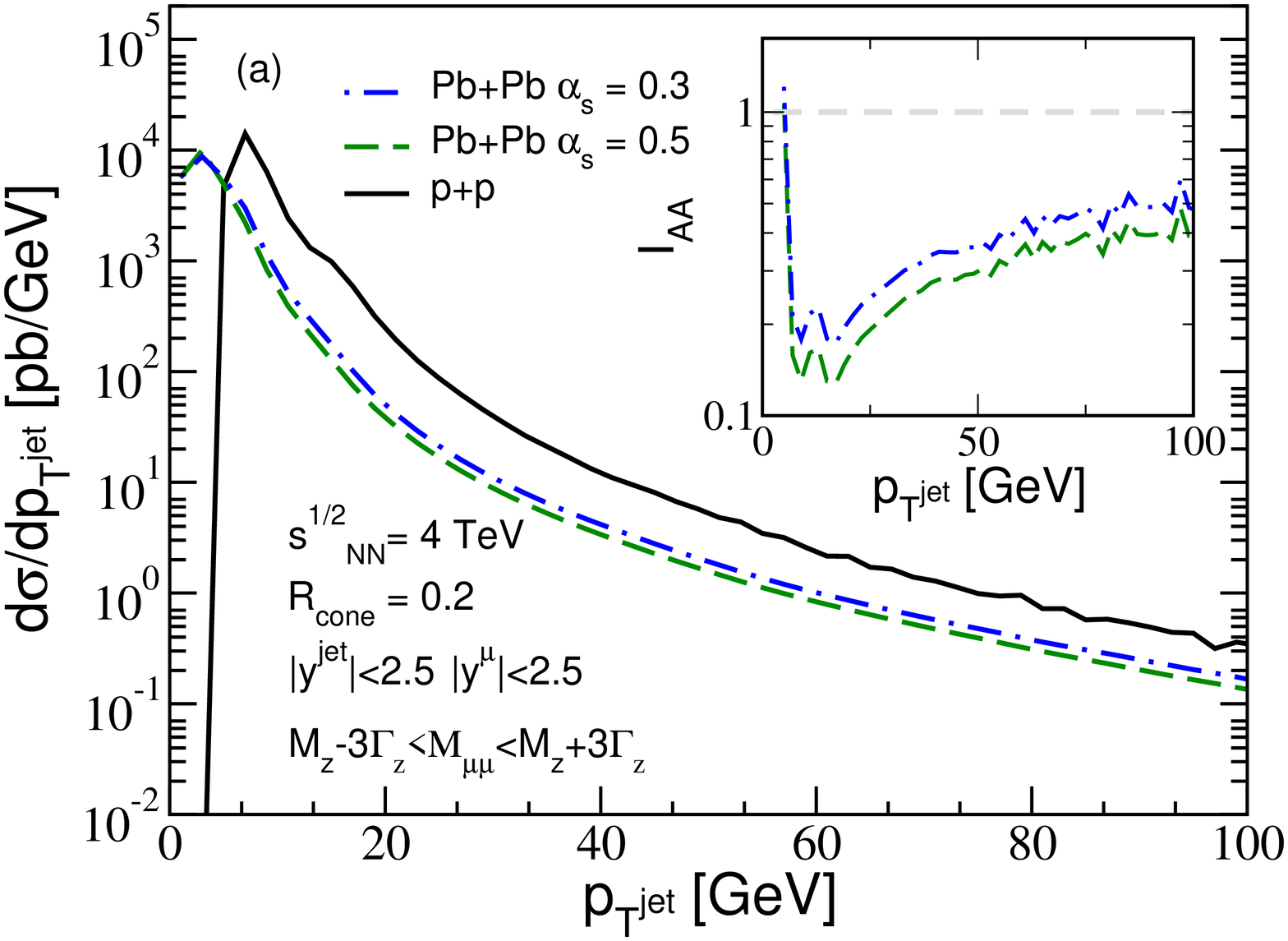}
}\vskip0.06\linewidth
\centerline{
\includegraphics[width=0.45\textwidth]{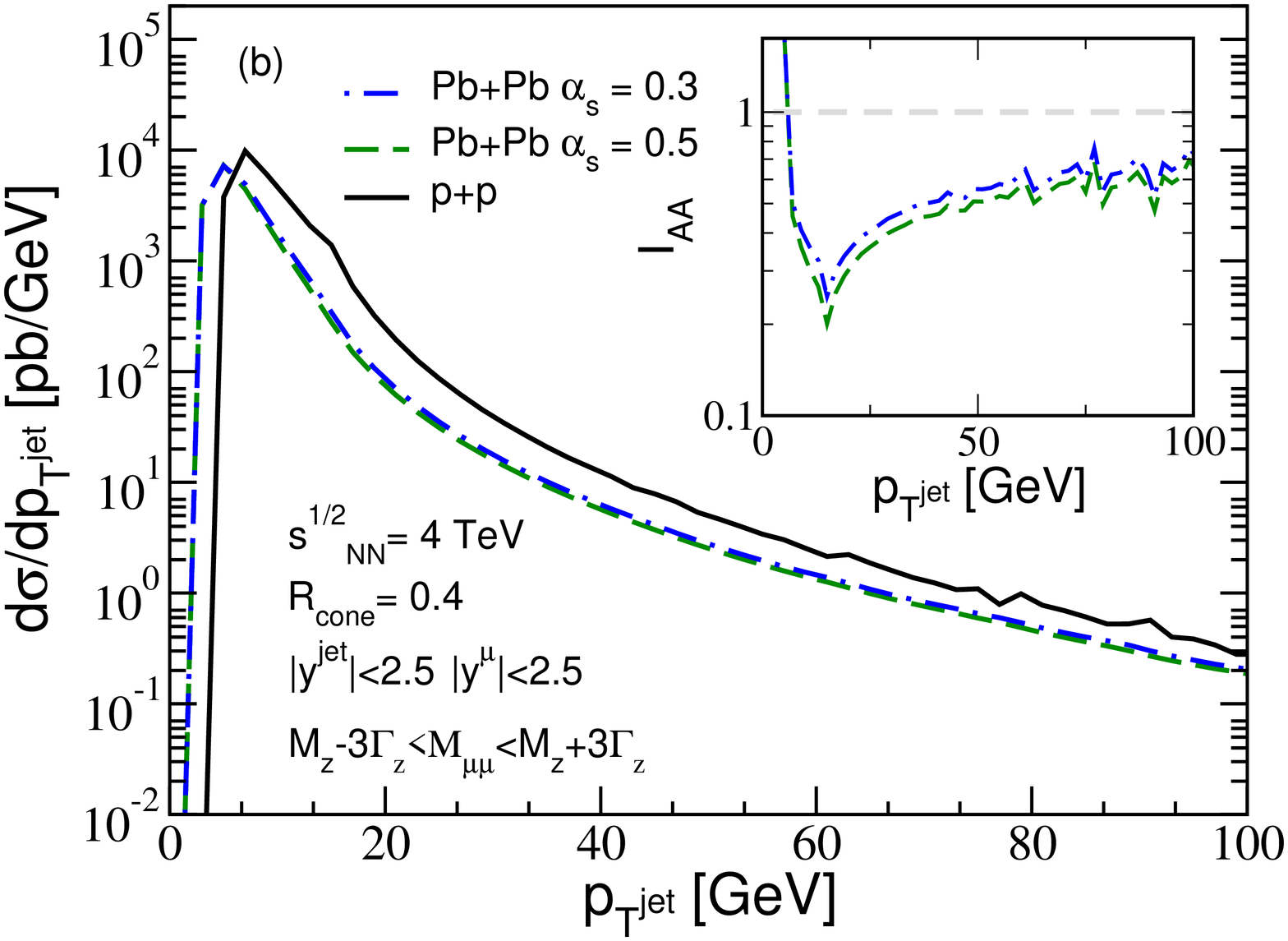}
}\vskip0.06\linewidth
\centerline{
\includegraphics[width=0.45\textwidth]{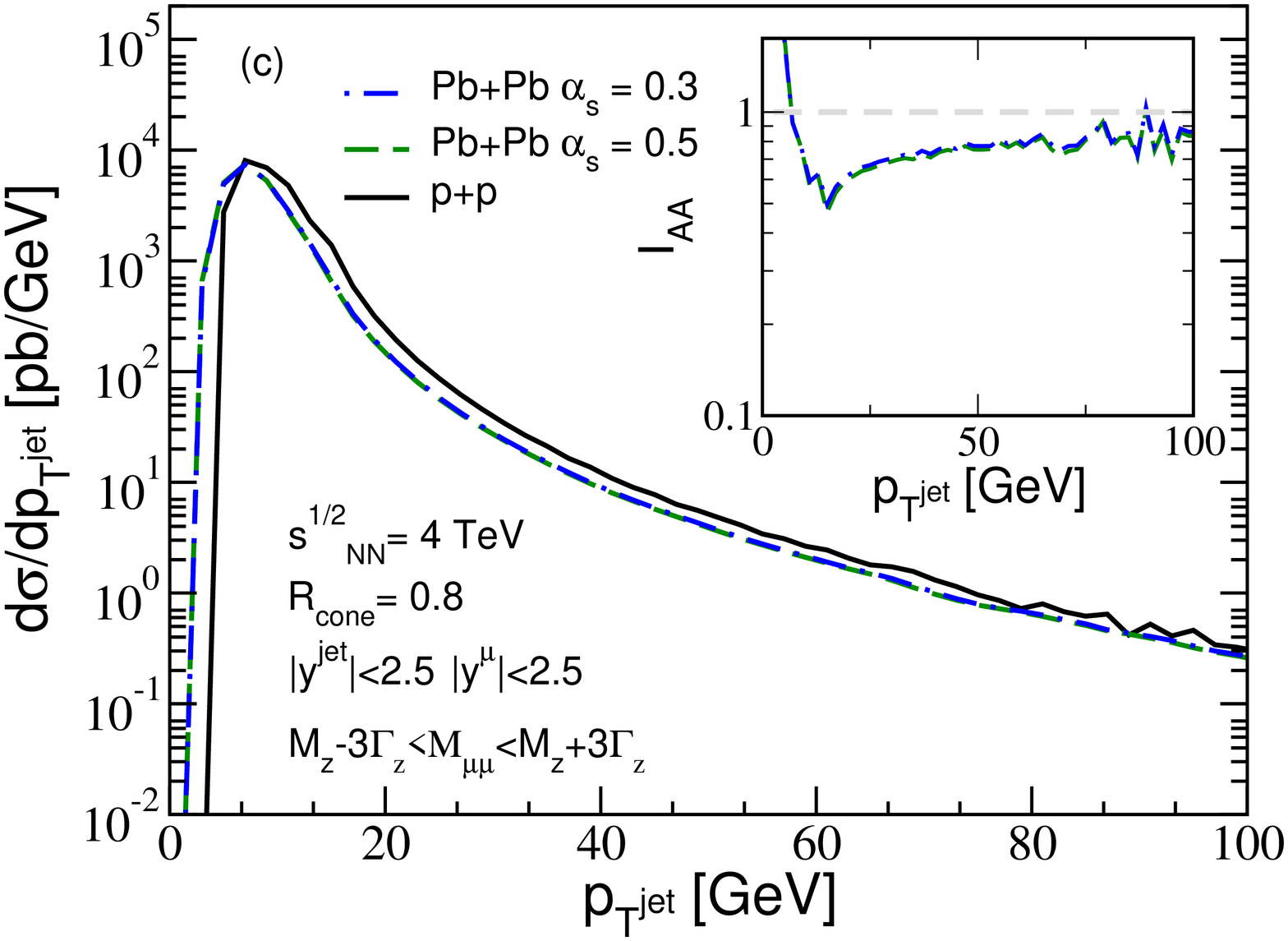}
}
\caption{(Color online)  The NLO $p_T$-differential cross section per nucleon pair for jets tagged with
$Z$/$\gamma^*\rightarrow \mu^++\mu^-$ in p+p and central Pb+Pb reactions. The tagging $Z$/$\gamma^*$ is required to have
$7 \; {\rm GeV} < p_T < 13 \; {\rm GeV}$.  Results are shown for three different values of jet cone radius $R = 0.2$ (upper panel) $R = 0.4$ (middle panel) and $R = 0.8$ (lower panel).
The ratio of the tagged QGP-modified cross section per nucleon pair to that in p+p, $I_{AA}$, is shown in the inset of each plot.}
\label{7_13_tagging}
\end{figure}

\begin{figure}
\centerline{
\includegraphics[width=0.45\textwidth]{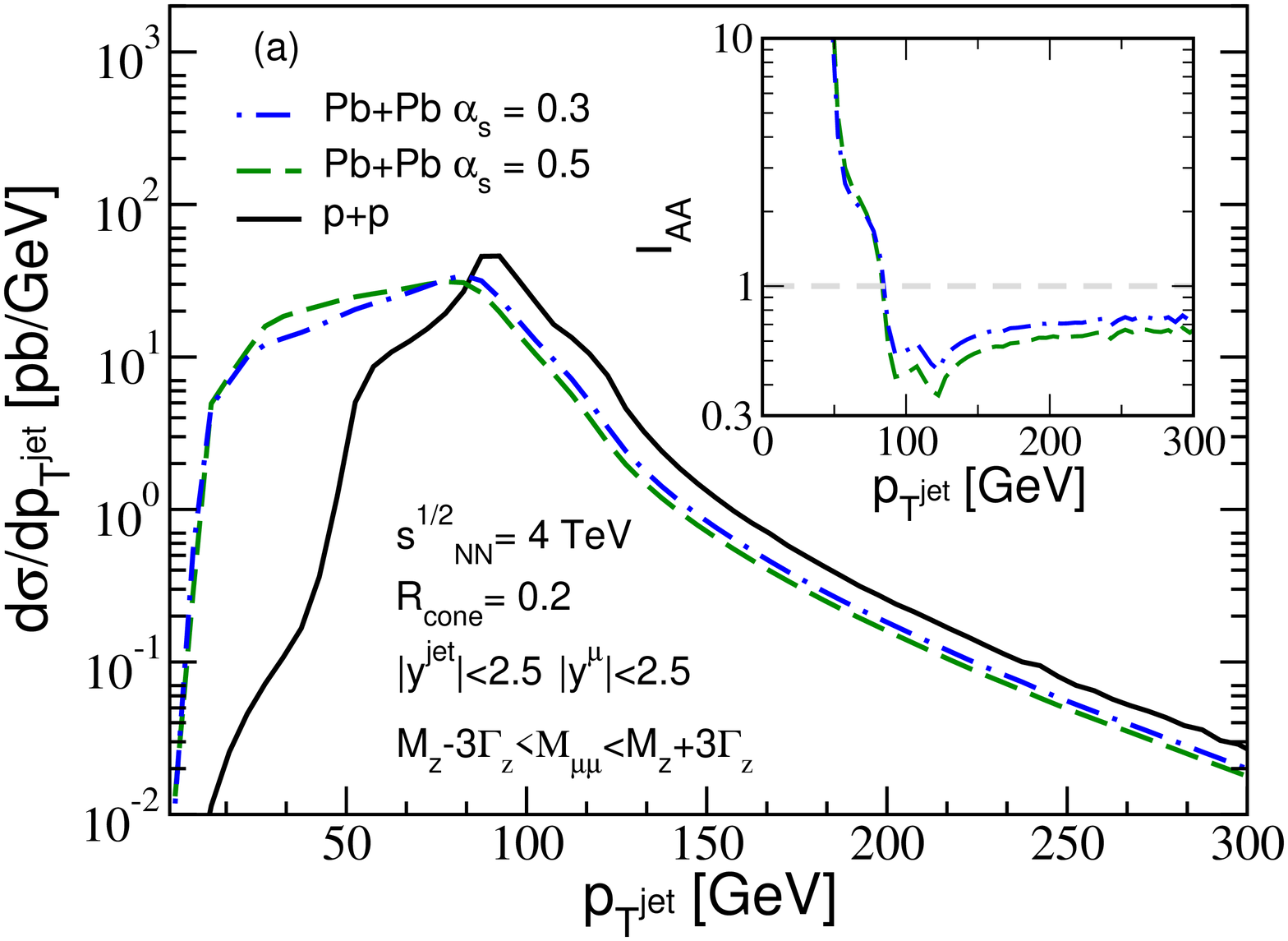}
}\vskip0.06\linewidth
\centerline{
\includegraphics[width=0.45\textwidth]{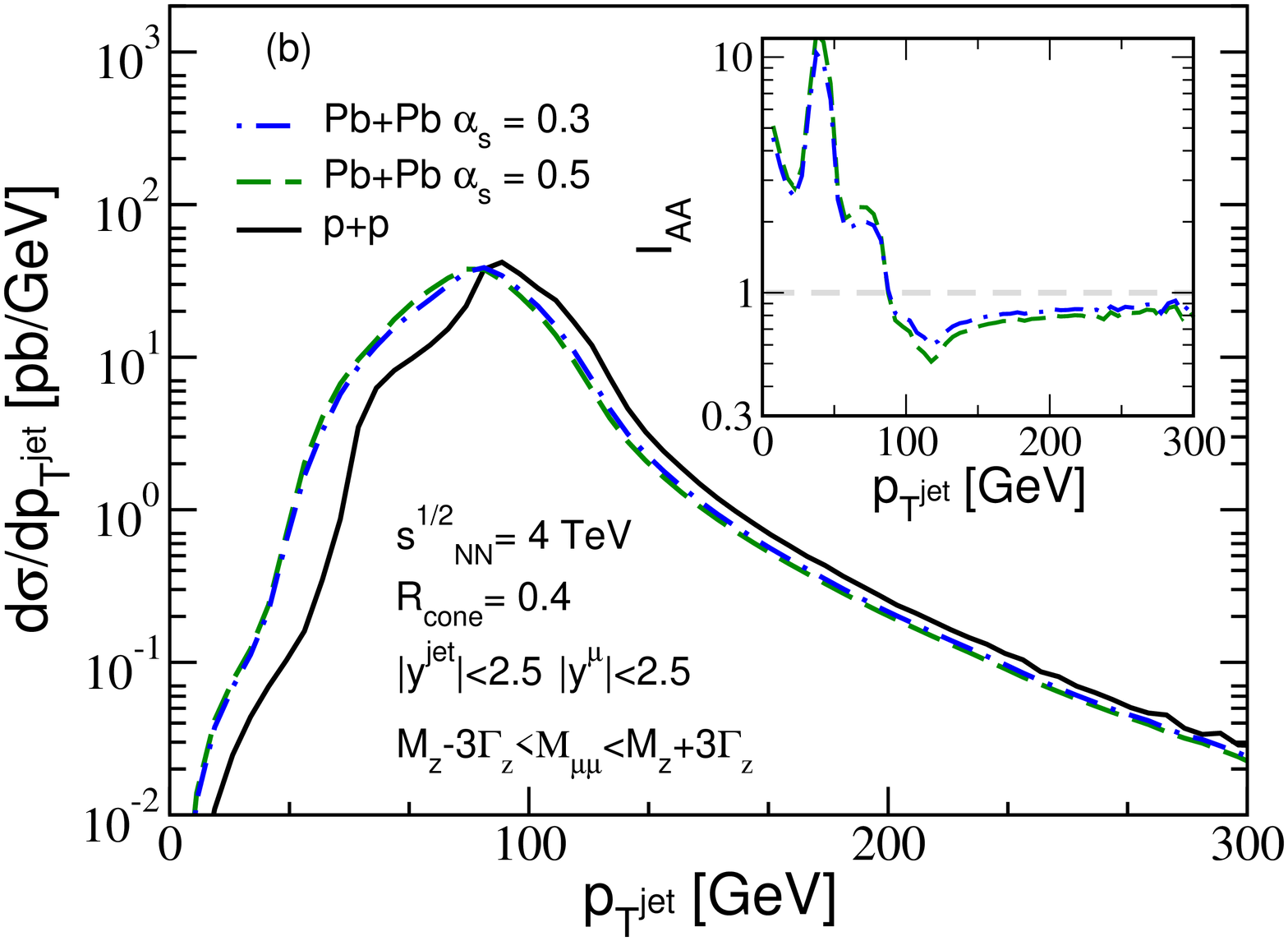}
}\vskip0.06\linewidth
\centerline{
\includegraphics[width=0.45\textwidth]{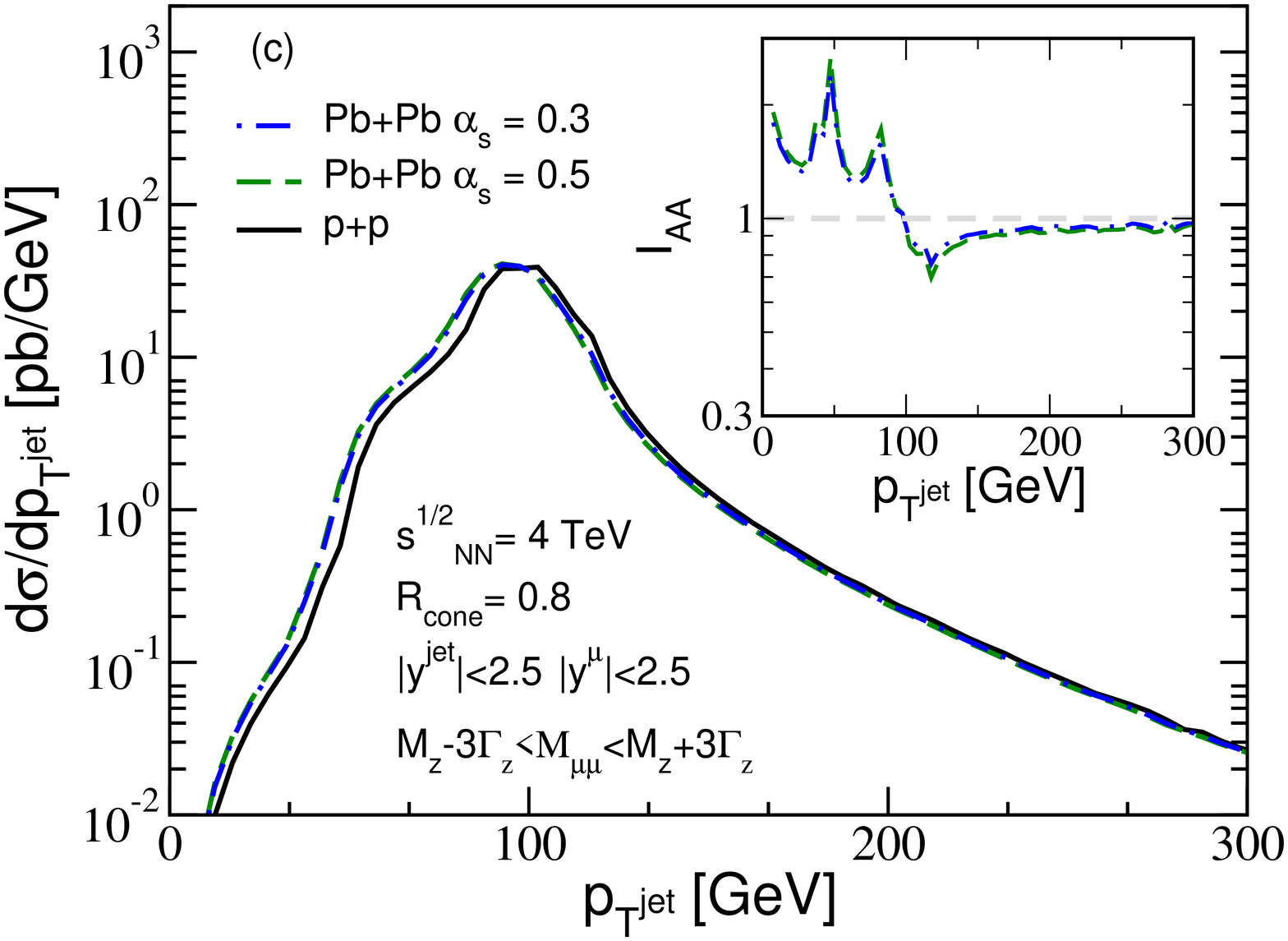}
}
\caption{(Color online) The same cross sections and ratios are shown as in Figure \ref{7_13_tagging} but now the tagging $Z$/$\gamma^*$
is required to have $92.5 \; {\rm GeV} < p_T < 112.5\; {\rm GeV} $.  Notice the sensitivity to the jet cone radius as $R$ varies from 0.2
to 0.8 (upper to lower panels) and also the significant enhancement in Pb+Pb relative to p+p for $p_{T\,(\rm jet)} \lesssim 100$
GeV.}
\label{925_1125_tagging}
\end{figure}

The NLO $p_T$-differential jet cross section with a tagging $Z$/$\gamma^*$ required to have $7 \; {\rm GeV} < p_T < 13\; {\rm GeV}$ is shown in Figure~\ref{7_13_tagging}.
This is the first order in perturbation theory at which the distribution of jets relative to the constrained $Z$/$\gamma^*$ momentum can be studied.
We present simulations for three different values of the jet cone radius, $R = 0.2, 0.4, 0.8$, and two different values for the jet-medium coupling strength, $\alpha_s = 0.3, 0.5$.
The main physics effect illustrated by this figure is the QGP-induced modification to the vacuum parton shower. Specifically, its broadening
\cite{Vitev:2008rz,Vitev:2005yg} implies that part of the jet energy is redistributed outside of the jet cone and the differential jet distribution is downshifted
toward smaller transverse momenta. The smaller the jet cone radius the more pronounced this effect is, as is clearly seen in the panels of Figure~\ref{7_13_tagging}.
As the jet cone radius is increased, the medium-modified curves approach the p+p result, as more and more of the medium-induced bremsstrahlung is recovered in the jet.
The variation in the magnitude of $I_{AA}^{\rm jet}$ at transverse momenta larger than the $p_T$ of the tag is controlled by the shape of the jet spectrum.
The most striking feature is the sharp transition from tagged jet suppression above $p_{T\, (Z)}$  to tagged jet enhancement
below $p_{T\, (Z)}$. This transition is a unique prediction of jet quenching for tagged jets and will be experimental evidence for strong final-state interactions
and parton energy loss in the QGP. Note that while the same energy redistribution occurs for inclusive jets~\cite{Vitev:2009rd}, the monotonically
falling spectrum prevents the observation of such enhancement in inclusive jet measurements.

Next, we consider the same cross sections and ratios for the tagging $Z^0$/$\gamma^*$ now required to have $92.5 \; {\rm GeV} < p_T < 112.5\; {\rm GeV} $.
Our results are shown in Figure~\ref{925_1125_tagging}.  The cross section enhancement in Pb+Pb relative to p+p for $p_{T\,(\rm jet)} \lesssim 100$ GeV
is quite dramatic. This is particularly true for the smallest cone radius, $R = 0.2$, where the medium modified result may become an order of
magnitude larger than in p+p, see $I_{AA}$ in the inset of the upper panel. The fluctuations in $I_{AA}^{\rm jet}$ arise from the
numerical accuracy of the simulation and the rigid acceptance cuts for both decay muons. In practice, these fluctuations will be smeared out.
 We conclude that a transverse momentum for the $Z^0/\gamma^*$ tag on the
order of $p_T \sim m_Z$ will be optimal to establish the QGP effects on jet propagation and distortion in heavy ion reactions if sufficient integrated
luminosity becomes available.

We finally present results for jets tagged with $Z^0$/$\gamma^*\rightarrow \mu^++\mu^-$ without imposing restrictions on the
transverse momentum of the tagging particle. This is precisely the way in which the Tevatron II measurements were done~\cite{Abazov:2008ez}.
The resulting cross sections are essentially cross sections for a special subset of processes that contribute to inclusive jet
production. The alteration of these ``inclusive'' cross sections in high-energy nucleus-nucleus collisions can be studied through the nuclear modification ratio:
\begin{equation}
R_{AA}^{\rm jet}(R,\omega_{\min}) = \frac{1}{\langle N_{\rm bin} \rangle}\frac{d\sigma_{AA}}{d p_{T\,(Q)}}
\Bigg / \frac{d\sigma_{pp}}{ d p_{T\,(\rm jet)}} \; .
\label{raa}
\end{equation}
\begin{figure}
\vskip0.035\linewidth
\centerline{
\includegraphics[width=0.45\textwidth]{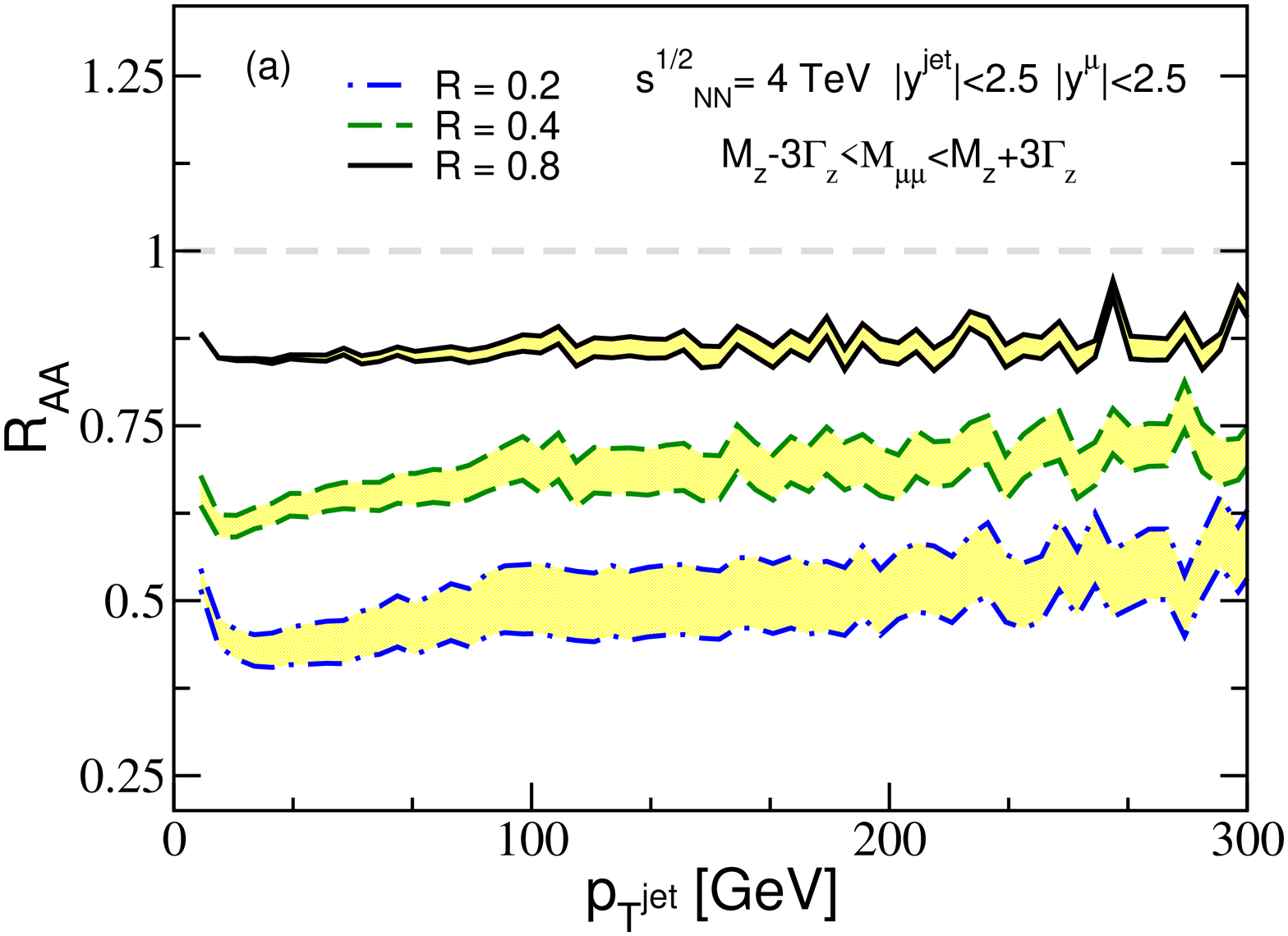}}
\vskip0.1 \linewidth
\centerline{
\includegraphics[height=0.33\textwidth,width=0.45\textwidth]{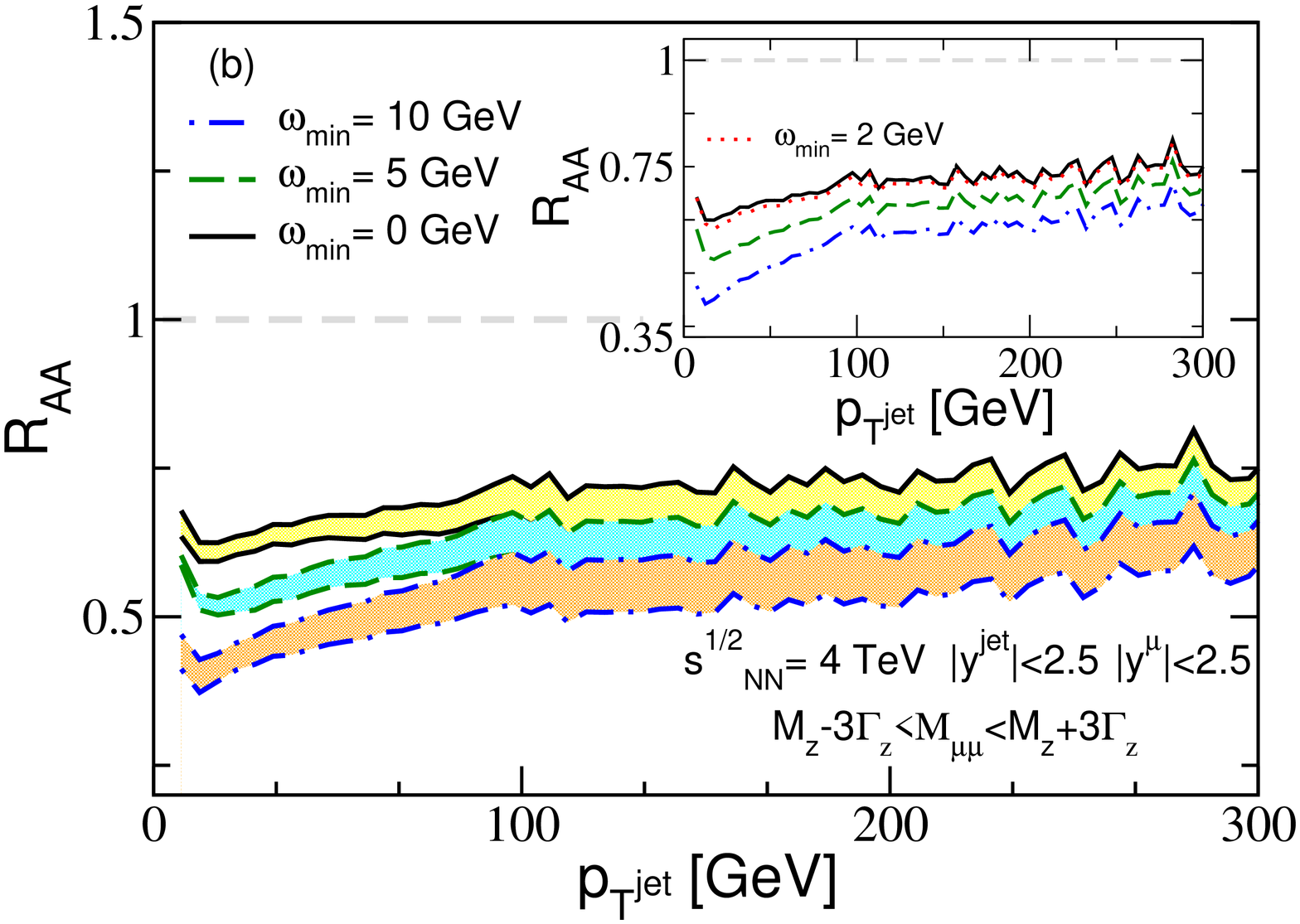}
}
\caption{(Color online) NLO results for jets tagged with $Z$/$\gamma^*\rightarrow \mu^++\mu^-$ presented as the ratio of the
QGP-modified cross section per nucleon pair to that in p+p. Note that there is no $p_T$ restriction on the $Z$/$\gamma^*$.
In the top panel we fix $\omega_{\min} = 0$ and vary $R$, whereas in the bottom panel we fix $R = 0.4$ and vary $\omega_{\min}$.
The insert shows that the variation of $R_{AA}(\omega_{\min})$ for $\omega_{\min} < 5$~GeV is negligible.}
\label{inclusivity}
\end{figure}
The NLO $p_T$-differential result for $R_{AA}$ is shown in Figure \ref{inclusivity}.  In the top panel we fixed $\omega_{\min} = 0$~GeV and varied $R$.
Bands represent a variation in the coupling strength of the jet parent parton to the QGP medium between $\alpha_s=0.3$ and  $\alpha_s=0.5$.
The strong dependence on the jet radius $R$ emphasizes the pronounced difference in the angular distribution of vacuum and medium-induced
parton showers in the framework of the GLV approach to parton energy loss in the QGP~\cite{Vitev:2005yg}.
In the  bottom panel of Figure \ref{inclusivity} we fixed $R = 0.4$ and varied $\omega_{\min}$. The observed additional suppression in $R_{AA}^{\rm jet}$
reflects the additional energy lost through the elimination of  the medium-induced semi-hard partons of $p_T < \omega_{\rm min}$ from  the jet definition.
The insert shows that there is
little difference in the simulated $R_{AA}^{\rm jet}$ for $\omega_{\min}=0$~GeV and $\omega_{\min}=2$~GeV and, consequently, that the medium-induced gluon spectrum
in the heavy ion reactions at the LHC are typically of larger energy ($\omega\sim 5$~GeV).

\section{Conclusions}\label{results}

Pb+Pb collisions at the LHC  at center-of-mass energies up to 5.5 TeV per nucleon pair  are expected to consolidate
the evidence for the creation of the QGP in ultrarelativistic  heavy ion reactions. Such collisions will also probe the
plasma properties  in a new temperature and energy density regimes and open final-state channels for jet tomography~\cite{Vitev:2009rd}
of the QGP that are currently inaccessible to experiment or limited by statistics.  One of these channels, and the
focus of the work presented in this paper, is the $Z^0/\gamma^*$+jet measured through the  $Z^0/\gamma^* \rightarrow
\mu^++\mu^-$ dilepton decay channel. Because of the short $Z^0$ production time and the fact that the decay dileptons
reach the detectors unscathed by the strong interactions, this channel has been proposed as a way to experimentally
constrain precisely the initial energy of the associated jet and through comparison to theory quantify the jet quenching
properties of the QGP \cite{Srivastava:2002kg,Kunde:1900zz,Lokhtin:2004zb}.

In this paper we examined the tagging power of the $Z^0/\gamma^*$ in p+p and A+A collisions at leading and next-to-leading orders. The main advantage of the NLO calculation is the ability to precisely predict the transverse momentum distribution of jets associated with a dimuon tag in a narrow $p_T$ interval. We found that, in contrast to the naive tree-level result, to ${\cal O}(G_F\alpha_s^2)$ there is very large $ \sim \pm 25\% $ uncertainty in the jet transverse energy distribution relative to the dimuon tag $p_T$. Our results demonstrate that direct experimental determination of the jet fragment distribution $p_{T\,(\rm h)}/p_{T\,(\rm jet)}$  without  adequate NLO theory is not realistic.  Some of the variation in the transverse momentum of the jet is likely to be eliminated by additional kinematic cuts but at the cost of reducing the already small cross section.  We conclude that it is not clear whether jet tagging with $Z^0/\gamma^*$ at the expense of a substantially reduced production rate offers significant advantages relative to direct photon tagging.

We also presented for the first time a calculation of tagged jet cross sections in heavy ion collisions up to next-to-leading order. Together with earlier tree-level and one-loop results for inclusive jets~\cite{Vitev:2008rz,Vitev:2009rd}, our work paves the way for more detailed and systematic theoretical and experimental studies of jet propagation in the background of strongly-interacting non-Abelian plasmas. It builds up a suite of differential jet observables to probe the microphysics behind the in-medium modification of parton showers. This new direction of research is promising in that it has the potential to overcome the limitations of leading particles~\cite{Bass:2008rv} and leading particle
correlations as tomographic probes of the QGP. Comparison between jet theory and upcoming experimental measurements at RHIC and at the LHC will help constrain theoretical model assumptions and approximations to parton propagation in hot and dense nuclear matter.  Elimination of unreliable phenomenology will, in turn, dramatically reduce the systematic error bars on the extracted plasma properties.

As a first example of tagged jet cross sections in nuclear collisions we calculated $Z^0/\gamma^*$+jet  production process in $\sqrt{s_{NN}}=4$ TeV central Pb+Pb
reactions at the LHC. We demonstrated that without constraints on the momentum of the $Z^0/\gamma^*$ decay dimuons, the quenching of tagged jets as a function of the jet cone radius $R$ and momentum acceptance cut $\omega_{\rm min}$ exhibits all features characteristics of inclusive jet measurements. Specifically, the variable attenuation rate of jet production $R_{AA}^{jet}(R,\omega_{\min})$ provides first-hand information about the differential spectrum of the QGP-induced gluon bremsstrahlung. In contrast, $I_{AA}^{jet}(R,\omega_{\min})$ yields additional insights into the characteristics of the medium-induced parton showers.  We have found that direct determination of the mean number of medium-induced gluons \cite{Neufeld:2010sz} is hampered by NLO corrections.  However, the dramatic transition from suppression of tagged jets above the $Z^0/\gamma^*$ $p_T$ to strong enhancement below this reference transverse momentum characterizes the probabilistic distribution of the energy lost by energetic partons as they traverse the QGP medium. Experimental measurements of such jet enhancement will provide an unambiguous confirmation of the dominant role of final-state interactions in the experimentally observed modification of jet and particle production cross sections in nucleus-nucleus collisions relative to an independent superposition of nucleon-nucleon scatterings.

We conclude by emphasizing that one can also take advantage of tagged jet physics in heavy ion collisions by using alternative final states. In particular, direct photon-tagged jets are characterized by significantly larger cross sections when compared to $Z^0/\gamma^*$-tagged jets and are accessible in central Au+Au collisions at RHIC. It is a top
priority to provide accurate theoretical predictions~\cite{Vitev:2009rd} for these reactions and to study how the QGP responds to propagating hard partons~\cite{Neufeld:2009ep}
tagged by an electroweak boson. \\

{ \bf Acknowledgments:} We thank J. M. Campbell, R.~K.~Ellis and G. Hasketh for illuminating discussions, clarification of the D0 experimental acceptance for $Z^0/\gamma^*$+jet, and assistance with the implementation of the MCFM code. This research is  supported by the US Department of Energy, Office  of Science, under
Contract No. DE-AC52-06NA25396 and
in part by the LDRD program at LANL, by the Ministry of Education of China with the Program NCET-09-0411,
by National Natural Science Foundation of China with Project No. 11075062,
and CCNU with Project No. CCNU09A02001.\\

Note added in proof: We have investigated initial-state cold nuclear matter 
effects~\cite{Neufeld:2010dz} in the context of the upcoming CMS inclusive $Z^0$ 
measurements~\cite{prcomm} and fond them to be on the order of a few $\%$.

\vspace*{.3cm}

\begin{appendix}

\section{Lowest Order  cross section for $Z^0$/$\gamma^*$-tagged jets}
\label{lo}

\begin{figure}[t!]
\centerline{
\includegraphics[width = 3in]{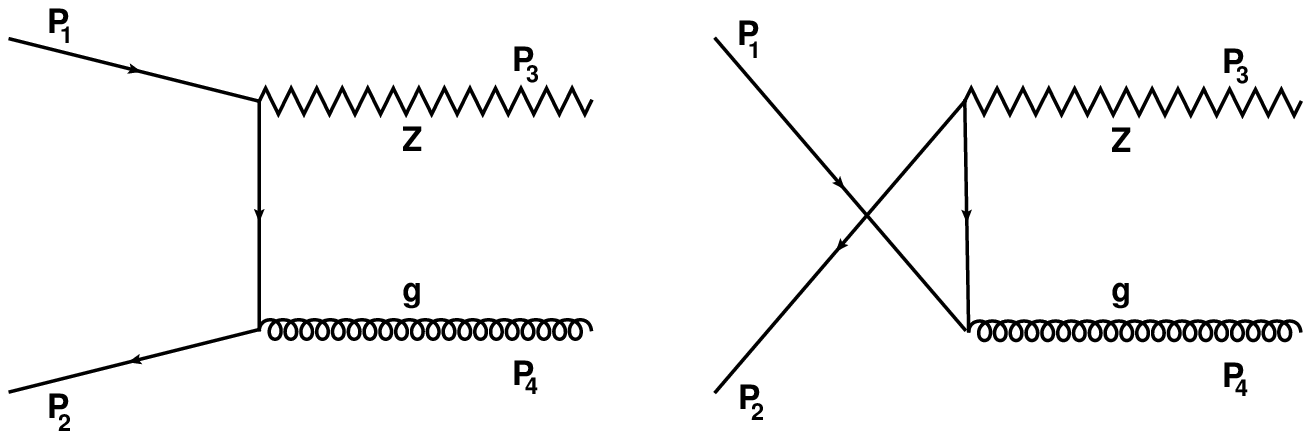}
}
\centerline{
\includegraphics[width = 3in]{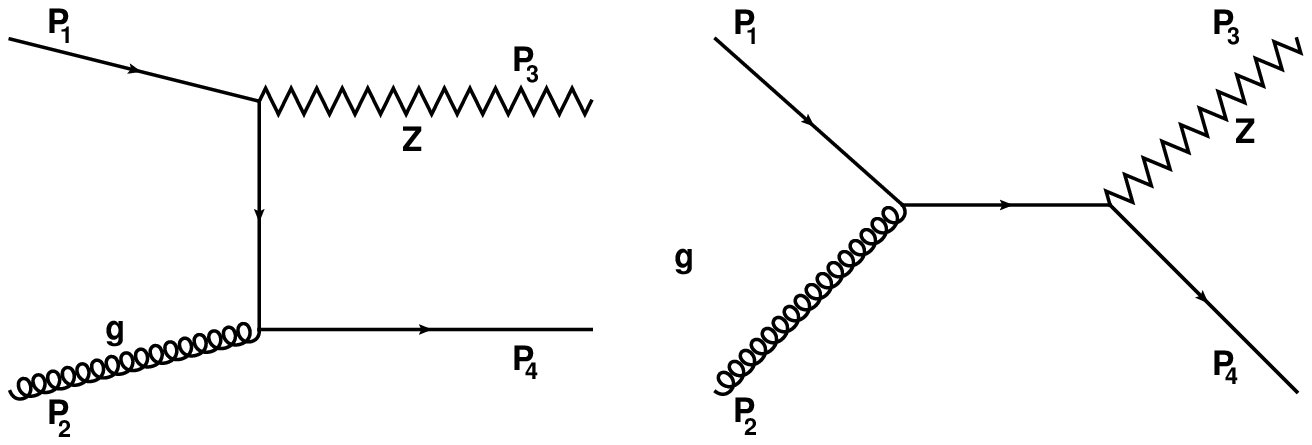}
}
\caption{LO diagrams contributing to the $Z^0$+jet production.  Here, $p_3$ is the momentum of the $Z^0$. At LO the contribution from
quark and gluon jets are conveniently separable; however, at NLO real gluon emission complicates that distinction.}
\label{quark_glue}
\end{figure}

The lowest order cross section for the $Z^0$/$\gamma^*$+jet production in hadronic collisions is well-known~\cite{Kajantie:1978qv}.
Nevertheless, it is useful to outline a few selected steps in its derivation in order to clarify the kinematics and production rate
for this process. To LO, only four diagrams, given in  Figure~\ref{quark_glue}, contribute. Our convention is to label the momenta
as given there.  Specifically, we have
\begin{eqnarray}
p_1 &=& (x_1 P_1 ,0,0, x_1 P_1)\;,  \nonumber \\
p_2 &=& (x_2 P_2 ,0,0, - x_2 P_2)\;, \nonumber \\
p_3 &=& (m_{T} \cosh y_3, \bald{p}_T, m_{T} \sinh y_3)\;, \nonumber \\
p_4 &=& (p_{T} \cosh y_4, -\bald{p}_T, p_T \sinh y_4) \nonumber \;,
\end{eqnarray}
where in all cases $p_3$ refers to the $Z^0$ and $p_4$ to the final-state quark or gluon jet.  Additionally, $P_1 =
P_2 = \sqrt{S}/2$,  $x_1, \; x_2$ are the momentum fractions carried by the initial-state partons, and $m_T = \sqrt{p_T^2 + m_Z^2}$.

The amplitude  for gluon jet production (upper panel of Figure~\ref{quark_glue}) is the sum of two matrix elements $M_{g1} + M_{g2}$.
Specifically, for the $q+\bar{q}\rightarrow Z^0+g$ process we have:
\begin{widetext}
\begin{eqnarray}
i M_{g1} &=& i  \frac{g_s g_z}{\sqrt{2} } \bar{v}(p_2,s_2)
(\gamma^\nu T^a)\frac{\slashed{p}_a}{p_a^2}(\gamma^\mu )
\left(R_q\left(1+\gamma^5\right) + L_q(1-\gamma^5)\right)u(p_1,s_1)\epsilon^*_\nu \epsilon^*_\mu  \;, \\
i M_{g2} &=& -i  \frac{g_s g_z}{\sqrt{2} }\bar{v}(p_2,s_2)(\gamma^\mu )
 \left(R_q\left(1+\gamma^5\right) + L_q(1-\gamma^5)\right)
\frac{ \slashed{p}_b}{p_b^2} (\gamma^\nu T^a) u(p_1,s_1)\epsilon^*_\nu \epsilon^*_\mu \;,
\end{eqnarray}
where $p_a = p_2 - p_4$ and $p_b = p_1 - p_4$.  The contribution for quark jet production (lower panel of Figure~\ref{quark_glue})
is the sum of $M_{q1} + M_{q2}$ for the process $q+g \rightarrow Z^0+q$:
\begin{eqnarray}
i M_{q_1} &=& i  \frac{g_s g_z}{\sqrt{2} }\bar{u}(p_4,s_4)(\gamma^\nu T^a)
\frac{i \slashed{p}_a}{p_a^2}(\gamma^\mu ) \left(R_q\left(1+\gamma^5\right) + L_q(1-\gamma^5)\right)u(p_1,s_1)
\epsilon_\nu \epsilon^*_\mu  \;, \\
i M_{q_2} &=& -i  \frac{g_s g_z}{\sqrt{2} } \bar{u}(p_4,s_4)(\gamma^\mu )
\left(R_q\left(1+\gamma^5\right) + L_q(1-\gamma^5)\right)\frac{\slashed{p}_c}{p_c^2}
(\gamma^\nu T^a) u( p_1,s_1)\epsilon_\nu \epsilon^*_\mu  \;,
\end{eqnarray}
\end{widetext}
where  $p_c = p_1 + p_2$. In what follows, we use  the notation of Ref.~\cite{Field:1989uq}  and have defined:
\begin{equation}
g_z^2 = \frac{\pi \alpha}{2 x_w(1-x_w)},
\end{equation}
with $x_w \equiv \sin^2\theta_w \approx 0.223$ for the on-shell Weinberg angle  and $\alpha \approx 1/137.04$ \cite{pdg}.  Also, we have introduced:
\begin{eqnarray}
&&R_q^2 = 4 \, e_q^2 \, \sin^4 \theta_w\;, \nonumber \\
&&L_q^2 = \tau_q^2 - 4 \, e_q \, \tau_q \, \sin^2 \theta_w + 4 \, e_q^2 \, \sin^4 \theta_w \;, \nonumber
\end{eqnarray}
where $\tau_q$ is the weak isospin of quark $q$ (that is, $\tau = 1$ for $u,c,t$ and $\tau = -1$ for $d,s,b$) and $e_q$
is the fractional electric charge of quark $q$.

In obtaining  the  squared amplitudes, $|M_{g}|^2 = |M_{g1} + M_{g2}|^2$ and $|M_{q}|^2 = |M_{q1} + M_{q2}|^2$,
we perform the external gluon polarization sums  with the replacement $\sum_{\mu\nu} \epsilon^*_\mu \epsilon_\nu
\rightarrow -g_{\mu\nu}$. For massive vector bosons these sums read:
\begin{equation}
\sum_{\mu\nu}  \epsilon^*_\mu \epsilon_\nu  \rightarrow - \left(g_{\mu\nu} - \frac{(p_3)_\mu (p_3)_\nu}{m_Z^2}\right) \;.
\end{equation}
After straightforward color and Dirac algebra manipulations the final tree-level result is obtained as:
\begin{eqnarray}
|M_g|^2 &=& \frac{16  \pi^2 \, \alpha_s \, \alpha\,(R_q^2 + L_q^2) }{9 x_w(1-x_w)}
\left(\frac{\hat{t}^2 + \hat{u}^2 + 2\,\hat{s}\,m_Z^2}{\hat{t} \hat{u}}\right) \;, \; \qquad  \label{Mg} \\
|M_q|^2 &=& - \frac{2  \pi^2 \, \alpha_s \, \alpha\,(R_q^2 + L_q^2) }{3 x_w(1-x_w)}
 \left(\frac{\hat{s}^2 + \hat{t}^2 + 2 \, \hat{u} \,m_Z^2}{\hat{s}\,\hat{t}}\right) \;, \; \qquad  \label{Mq}
\end{eqnarray}
where  $\alpha_s = g_s^2/4\pi$ and the Lorentz-invariant Mandelstaam variables are:
\begin{eqnarray*}
&& \hat{s} = (p_1 + p_2)^2\;, \;
\hat{t} = (p_1 - p_3)^2 \;, \;
\hat{u} = (p_2 - p_3)^2\; .
\end{eqnarray*}
The $\gamma^*$-tagged jet matrix elements squared  can be obtained from Eqs. ~(\ref{Mg}) and (\ref{Mq})   with the replacement:
\eqf{
\frac{(R_q^2 + L_q^2)}{x_w(1-x_w)}\rightarrow 2 e_q^2 \; , 
}
as well as letting $m_Z^2 \rightarrow Q^2$, where $Q$ is the photon virtuality. While in p+p reactions
we are interested in the total jet rate production rate,  $ \propto |M_g|^2 + |M_q|^2 + |M_{\bar{q}}|^2 $
($|M_{\bar{q}}|^2 =|M_{{q}}|^2$), the separation into final-state quark and gluon jets will prove useful
in the treatment of the medium-induced radiative energy loss in A+A reactions. Note that when
the initial-state partons are interchanged the matrix elements can be trivially obtained with the
substitution $\hat{t} \leftrightarrow \hat{u}$.

In the collinear factorization approach the differential $Z^0/\gamma^*$+jet cross section is 
evaluated as follows:
\eqf{
d \sigma &= \sum_{g,q,\bar{q}}  \int dx_1 dx_2 \; f(x_1,Q) f(x_2,Q) \frac{1}{2 x_1 x_2 S} |M|^2  \\
& \!\!\!\!\!  \times (2\pi)^4 \delta^4(p_1 + p_2 - p_3 - p_4)
\frac{d^3 p_3}{2E_3(2\pi)^3}  \frac{d^3 p_4}{2E_4(2\pi)^3}  \;.
}
Further, one notices that $d^3 p_i = E_i \, d y_i \, d^2 p_{T_i}$ and
$$\delta^4(p_1 + p_2 - p_3 - p_4) = \frac{2}{S}\delta(x_1 - \bar{x}_1)
\delta(x_2 - \bar{x}_2)\delta^2(\bald{p}_{T_3} - \bald{p}_{T_4})\;, $$
where $\bar{x}_1$ and $\bar{x}_2$ are the solutions to the coupled equations:
\begin{eqnarray*}
(x_1 + x_2)\sqrt{S}/2 - m_{T_3} \cosh y_3 - p_{T_4} \cosh y_4 &=& 0 \; ,\\
(x_1 - x_2)\sqrt{S}/2 - m_{T_3} \sinh y_3 - p_{T_4} \sinh y_4 &=& 0\; .
\end{eqnarray*}
We find:
\begin{eqnarray}
\bar{x}_1 &=& \frac{m_{T_3} \, e^{y_3} + p_{T_4} \, e^{y_4}}{\sqrt{S}} \;, \\
\bar{x}_2 &=& \frac{m_{T_3} \, e^{-y_3} + p_{T_4} \, e^{-y_4}}{\sqrt{S}}\; .
\end{eqnarray}
and  LO cross section is then given by ($p_T=p_{T_3}=p_{T_4}$)
 \begin{equation}
 \frac{d \sigma}{d y_3 \, d y_4 \, d^2 p_T} = \sum_{g,q,\bar{q}} f(\bar{x}_1)f(\bar{x}_2)
 \frac{|M|^2}{(2\pi)^2 \, 4 \, \bar{x}_1 \, \bar{x}_2 \, S^2} \; .
\end{equation}

\section{ The $Z/\gamma^*\rightarrow l^++l^-$ Dalitz Decay}
\label{dalitz}

In this Appendix we discuss the  $Z^0/\gamma*$ decay to dileptons with the goal of addressing the
experimental measurements at the Tevatron and at the LHC. Our approach is to work in the rest
frame of the parent particle. If this
particle has mass  $m_Z$, its differential decay rate is given by:
\eqf{
\frac{d\Gamma}{d\Omega} = \frac{1}{32\pi^2}|M|^2\frac{|\bald{p}|}{m_Z^2}\;,
}
where $\bald{p}$ is the momentum of one of the final-state  particles
(the other particle has momentum $-\bald{p}$) and $|M|^2$
is the matrix element squared for this decay.   For instance, in the case
of $\gamma^*\rightarrow \mu^++\mu^-$:
\eqf{
i M &= -i e \bar{u}(p,s_2) \gamma^\nu v(p,s_1)\epsilon^*_\nu  \\
& \Rightarrow \;\; \frac{1}{3}\sum_{spin} |M|^2 = \frac{4 e^2 Q^2}{3} \; ,
}
where $Q$ is the virtuality of the photon (that is, the invariant mass of the muon pair).
The calculation of $|M|^2$ for $Z \rightarrow \mu^++\mu^-$ is analogous:
\begin{widetext}
\eqf{
i M &= -i \frac{g_z}{\sqrt{2}}  \bar{u}(p,s_2) \gamma^\nu \left(R_e(1+\gamma^5)
+ L_e(1-\gamma^5)\right)v(p,s_1)\epsilon^*_\nu \;\; \Rightarrow  \;\; \frac{1}{3}\sum_{spin}
|M|^2 = \frac{4 g_z^2 m_Z^2(R_e^2 + L_e^2)}{3} \; ,
}
\end{widetext}
where $R_e = 2 x_w$ and $L_e = 2 x_w - 1$ (see Appendix~\ref{lo} for the definition of $x_w$ and $g_z$).
Clearly, the distribution of decay products is isotropic.

To evaluate the  cross section for the $Z^0/\gamma^*(\rightarrow \mu^++\mu^-)$+jet process
we use the experimentally determined branching ratio $\Gamma_{\mu\mu}/\Gamma \approx 0.037$ \cite{pdg}.
The velocity of the $Z^0$ boson in the lab frame, in terms of the variables discussed in Appendix \ref{lo},
is given by:
\eqf{
\vec{\beta} = \frac{\left(\bald{p}_T, m_{T} \sinh y_3\right)}{m_{T} \cosh y_3}\; .
}
In the $Z^0$ rest frame, the 4 momenta of the two decay muons are given by:
\begin{eqnarray*}
p_{1z} &=& \frac{m_Z}{2}\left(1,\sin\theta\cos\phi,\sin\theta\sin\phi,\cos\theta\right) \;,   \\
p_{2z} &=& \frac{m_Z}{2}\left(1,-\sin\theta\cos\phi,-\sin\theta\sin\phi,-\cos\theta\right) \;,
\end{eqnarray*}
where $\theta$ and $\phi$ are uniformly distributed over the solid angle.  The 4 momenta of the
two decay muons in the lab frame can be obtained with an appropriate Lorentz boost
$p_{L} = \hat{U} p_{z}$
where $\hat{U}$ is given by:
\[ \left( \begin{array}{cccc}
\gamma & \beta_x\gamma & \beta_y\gamma & \beta_z\gamma \\
\beta_x\gamma & 1+(\gamma-1)\frac{\beta_x^2}{\beta^2} & (\gamma-1)\frac{\beta_x \beta_y}{\beta^2} & (\gamma-1)\frac{\beta_x \beta_z}{\beta^2} \\
\beta_y\gamma & (\gamma-1)\frac{\beta_y \beta_x}{\beta^2} & 1+(\gamma-1)\frac{\beta_y^2}{\beta^2} & (\gamma-1)\frac{\beta_y \beta_z}{\beta^2} \\
\beta_z\gamma & (\gamma-1)\frac{\beta_z \beta_x}{\beta^2} & (\gamma-1)\frac{\beta_z \beta_y}{\beta^2} &1+(\gamma-1)\frac{\beta_z^2}{\beta^2} \end{array} \right) \;.  \] 
This approach enables one to determine the distribution of the final-state dileptons and to simulate any relevant experimental acceptance cuts.

\section{The role of quark versus gluon energy loss in inclusive and tagged jet quenching measurements  }
\label{qgvariation}

At next-to-leading order perturbative splitting processes [$q(\bar{q})\rightarrow q(\bar{q})g$, $q(\bar{q})\rightarrow gq(\bar{q})$,  $g\rightarrow gg$, $g\rightarrow q\bar{q}$] complicate the flavor identification of the jet parent parton. In proton-proton collisions the ambiguity of quark vs gluon jets primarily affects flavor-tagged jet observables~\cite{Banfi:2006hf}. In nucleus-nucleus collisions the primary concern is the possible uncertainty in the amount of medium-induced gluon bremsstrahlung, which is proportional to the average squared color charge ($C_F=4/3$ and $C_A=3$ for quarks and gluons, respectively). When parton energy loss is not very small relative to the initial jet energy, the factor of $C_A/C_F=2.25$ difference between $\Delta E_q$ and  $\Delta E_g$ is greatly reduced~\cite{Vitev:2008jh}. It is nevertheless instructive to examine how changes in the relative abundances of quark and gluon jets (possibly arising in higher order calculations) affect the observable $R_{AA}^{\rm jet}$ and $I_{AA}^{\rm jet}$.

\begin{figure}[!t]
\vspace*{.2in}
\centerline{
\includegraphics[width = 3in,height=2.2in]{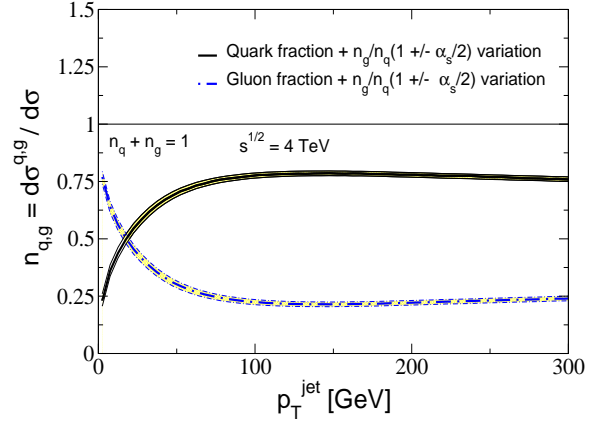}
}
\caption{ Quark and gluon jet fractions $n_q,\, n_g$ versus $p_T$ for $Z^0/\gamma^*$-tagged jets
in $\sqrt{s_{NN}}=4$~TeV p+p collisions at the LHC (evaluated at tree level). The band represents
a possible variation of $n_g/n_q$ of $\pm \alpha_s(m_{T\,Z})/2$.}
\label{qgvar}
\end{figure}

\begin{figure}[!b]
\centerline{
\includegraphics[width = 3in,height=2.2in]{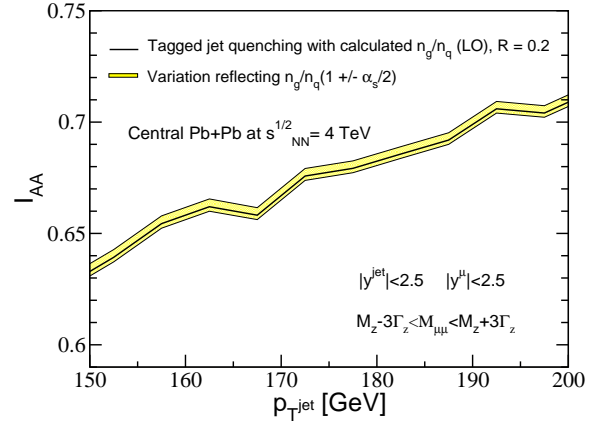}
}
\caption{ Effect of  $n_q,\, n_g$  variation on the suppression of the $Z^0/\gamma^* (\rightarrow \mu^++\mu^-)$-tagged
jet cross section in central Pb+Pb reactions at $\sqrt{s_{NN}}=4$~TeV at the LHC. We have chosen a
jet cone radius $R=0.2$. }
\label{IAAvar}
\end{figure}

A solution that we use for inclusive~\cite{Vitev:2009rd} and tagged jets is to evaluate at tree level
the quark and gluon jet fractions $n_q,\, n_g$  ($n_q+n_g = 1$) for the specific final state. In this
appendix we demonstrate that the sensitivity of inclusive and tagged jet suppression measures $R_{AA}$ and $I_{AA}$ to ${\cal O}(\alpha_s)$
corrections to this quark versus gluon jet mixing is small. Parton splitting is proportional to the strong
coupling constant and we focus on a possible variation in $n_g/n_q$ by this amount. Figure~\ref{qgvar}
shows the calculated quark and gluon jet fractions:
\begin{equation}
n_q = \frac{d\sigma^{q\; {\rm jet}}}{dp_T}\Bigg/\frac{d\sigma^{ {\rm jet}}}{dp_T}\;, \quad
n_g = \frac{d\sigma^{g\; {\rm jet}}}{dp_T}\Bigg/\frac{d\sigma^{ {\rm jet}}}{dp_T}\;,
\end{equation}
in $\sqrt{s_{NN}}=4$~TeV p+p collisions at the LHC. The detailed behavior of these fractions as a function of $p_T$,
including  the non-monotonic structure around $p_T \sim m_Z$, arises from the interplay of the
initial-state abundance of (anti)quarks and gluons and the behaviour of the short-distance scattering,
Eqs.~(\ref{Mg}) and (\ref{Mq}). Bands represent the following variation of the gluon to quark jet ratio:
 \begin{equation}
\frac{n_g}{n_q}  \rightarrow \frac{n_g}{n_q} \left( 1\pm \frac{\alpha_s(m_{T\,Z})}{2} \right) \; .
\end{equation}
Due to the smallness of strong coupling constant at large $p_T$, the size of the bands in Figure~\ref{qgvar} is also small.

The  effect of such $n_q/n_g$ uncertainty on $I_{AA}^{\rm jet}$ is shown in Figure~\ref{IAAvar}
for a dimuon tag  $92.5\; {\rm GeV} < p_T < 112.5\; {\rm GeV}  $ and a jet cone radius $R=0.2$.
Unlike in Figure~\ref{925_1125_tagging},  we have used a much smaller transverse momentum
range for the jet and a linear $I_{AA}^{\rm jet}$ scale to illustrate that the resulting uncertainty
in the quenching of $Z^0/\gamma^*$-tagged jets is of the order of $1\%$.

\begin{figure}[t!]
\centerline{
\includegraphics[width = 3in,height=2.1in]{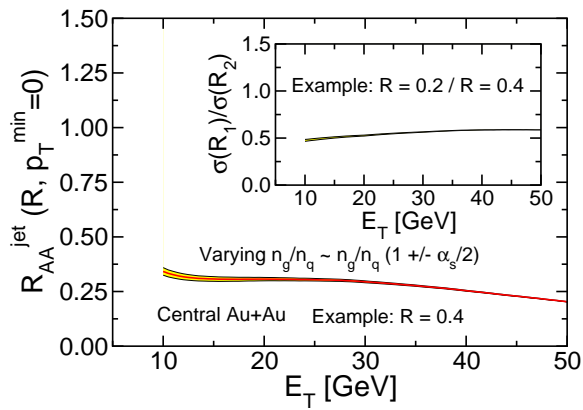}
}
\caption{ Effect of  $n_q,\, n_g$  variation on the suppression of the inclusive
jet cross section in central Au+Au reactions at $\sqrt{s_{NN}}=200$~GeV at RHIC.
We have chosen a jet radius $R=0.4$. Insert shows the same effect for the ratio of
two cross sections with $R=0.2$ and $R=0.4$, respectively. }
\label{RAAvar}
\end{figure}

Finally, we point out that for inclusive jets the variation in the cross section suppression
$R_{AA}^{\rm jet}$ is similarly small.  The  is illustrated in Figure~\ref{RAAvar} for RHIC energies
of $\sqrt{s_{NN}}=200$~GeV. Our calculation is for central Au+Au reactions with a choice for the
jet cone radius $R=0.4$~\cite{Vitev:2008jh}. Everywhere in the accessible $p_T$ range the uncertainty
on the attenuation of the jet rate in the final state is less than $5\%$.

\end{appendix}

\end{document}